\documentclass[
 reprint,
superscriptaddress,
nofootinbib,
 amsmath,amssymb,
prl
]{revtex4-2}

\usepackage{graphicx}
\usepackage{dcolumn}
\usepackage{bm}
\usepackage{braket}
\usepackage{mathtools}
\usepackage[dvipsnames]{xcolor}
\usepackage{hyperref}

\newcommand{\s}{\,\mathrm{s}}
\newcommand{\us}{\,\textrm{µs}}
\newcommand{\nm}{\,\mathrm{nm}}
  
\newcommand{\mm}{\,\mathrm{mm}}
\newcommand{\MHz}{\;\mathrm{MHz}}

\newcommand{\ba}{$^{137}\textrm{Ba}^+$\;}
\newcommand{\ground}{$\mathrm{S_{1/2}}\;$}
\newcommand{\groundm}{\mathrm{S_{1/2}}}

\newcommand{\pone}{$\mathrm{P_{1/2}}\;$}

\newcommand{\dthree}{$\mathrm{D_{3/2}}\;$}

\newcommand{\shelf}{$\mathrm{D_{5/2}}\;$}
\newcommand{\shelfm}{\mathrm{D_{5/2}}}

\begin{document}

\title{High-fidelity heralded quantum state preparation and measurement}

\author{A. S. Sotirova}
\email[Corresponding to: ]{ana.sotirova@oxionics.com}
\thanks{These authors contributed equally to this work.}
\affiliation{Department of Physics, University of Oxford, Oxford, OX1 3PU, UK}
\affiliation{Oxford Ionics, Oxford, OX5 1PF, UK}

\author{J. D. Leppard}
\thanks{These authors contributed equally to this work.}
\affiliation{Department of Physics, University of Oxford, Oxford, OX1 3PU, UK}

\author{A. Vazquez-Brennan}
\affiliation{Department of Physics, University of Oxford, Oxford, OX1 3PU, UK}

\author{S. M. Decoppet}
\affiliation{Department of Physics, University of Oxford, Oxford, OX1 3PU, UK}

\author{F. Pokorny}
\thanks{Now at Oxford Ionics Ltd.}
\affiliation{Department of Physics, University of Oxford, Oxford, OX1 3PU, UK}

\author{M. Malinowski}
\affiliation{Oxford Ionics, Oxford, OX5 1PF, UK}

\author{C. J. Ballance}
\affiliation{Department of Physics, University of Oxford, Oxford, OX1 3PU, UK}
\affiliation{Oxford Ionics, Oxford, OX5 1PF, UK}

\date{\today}

\begin{abstract}
We present a novel protocol for high-fidelity qubit state preparation and measurement (SPAM) that combines standard SPAM methods with a series of in-sequence measurements to detect and remove errors. The protocol can be applied in any quantum system with a long-lived (metastable) level and a means to detect population outside of this level without coupling to it. We demonstrate the use of the protocol for three different qubit encodings in a single trapped \ba ion. For all three, we achieve the lowest reported SPAM infidelities of $7(4) \times 10^{-6}$ (optical qubit), $5(4) \times 10^{-6}$ (metastable-level qubit), and $8(4) \times 10^{-6}$ (ground-level qubit).

\end{abstract}

\maketitle

\section{Introduction}
\label{sec:intro}
One of the requirements for building a quantum computer (QC) is the ability to perform qubit state preparation and measurement (SPAM) with low error \cite{divincenzo_physical_2000}. QC platforms based on atomic qubits allow for high-fidelity SPAM using optical pumping \cite{kastler_quelques_1950, brossel_greation_1952, hawkins_polarization_1953} and state-selective fluorescence \cite{acton_near-perfect_2006, nagourney_shelved_1986, sauter_observation_1986, bergquist_observation_1986}, avoiding the tradeoffs between SPAM fidelity and gate fidelity common for solid-state platforms \cite{purcell_resonance_1946, reed_fast_2010, walter_rapid_2017}. Following early work on high-fidelity SPAM in $^{40}\mathrm{Ca}^+$ \cite{myerson_high-fidelity_2008}, multiple techniques have been developed to extend these methods onto atomic qubits with more complex internal structure, such as ions with nuclear spin $I > 1/2$ \cite{an_high_2022, harty_high-fidelity_2014, leu_polarisation-insensitive_2024}, which can be advantageous for quantum computing, sensing, and timekeeping \cite{guggemos_frequency_2019, harty_high-fidelity_2014}. While previous experiments reported SPAM errors as low as $9.0(13)\times 10^{-5}$ \cite{an_high_2022}, high-fidelity qubit SPAM remains in general challenging and resource-intensive, with error sources including impure laser polarization, imperfect transfer pulses, and finite atomic lifetime.

In this work, we propose a novel protocol that alleviates many of these challenges, allowing for high-fidelity SPAM of a broad class of qubits in atomic and atom-like systems. We combine the standard SPAM methods with a number of mid-circuit non-demolition (QND) measurements \cite{grangier_quantum_1998, negnevitsky_repeated_2018} that allow us to detect and correct a broad range of errors. Specifically, these measurements are designed to raise a flag when a SPAM error occurs. We can use these flags to reject the corresponding experiment shot through post-selection  \cite{aaronson_quantum_2004}, repeat the cycle until no flag is raised (repeat-until-success) \cite{lim_repeat-until-success_2005} or incorporate them into the analysis of the algorithm's output \cite{nation_scalable_2021}. The protocol can be applied to different qubit encodings, such as optical (O), metastable (M), and ground (G) qubits in atomic systems \cite{allcock_omg_2021} and in any platform with a metastable manifold and a measurement that can detect population outside this manifold. Our protocol is closely connected to the idea of erasure conversion of leakage errors in quantum operations \cite{wu_erasure_2022, kubica_erasure_2022, kang_quantum_2023, varbanov_leakage_2020, yang_realizing_2022}.

We implement the protocol on three types of qubits in a single trapped \ba ion, achieving the lowest SPAM error reported to date, for any qubit, on all of them -- O: $7(4) \times 10^{-6}$, M: $5(4) \times 10^{-6}$, and G: $8(4) \times 10^{-6}$. We find that the in-sequence QND measurements reduce the SPAM error by over four orders of magnitude while only rejecting $< 10\%$ of the shots. These results demonstrate the versatility and error resilience of our SPAM protocol.

\section{SPAM Protocol}
\label{sec:spam-protocol}
\begin{figure*}[!ht]
    \includegraphics[width=\textwidth]{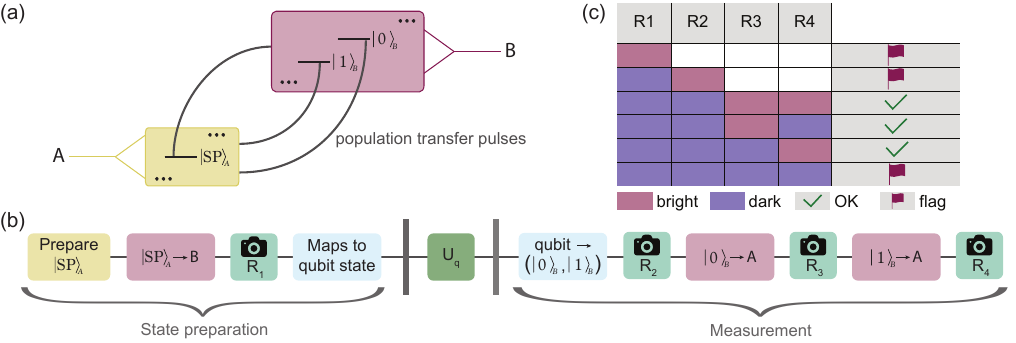}
    \caption{SPAM protocol. (a) Illustration of generic features of the quantum system capable of implementing the protocol. (b) The complete sequence of the SPAM protocol. The camera icons represent the population detection steps with corresponding results $R_1$--$R_4$. The steps shaded in light blue are only necessary if the qubit encoding $\left(\ket{0}, \ket{1}\right)$ is different from $\left(\ket{0}_B, \ket{1}_B\right)$. The step $U_q$ represents a generic quantum algorithm in which the qubit takes part. (c) A truth table showing whether or not an error flag is raised based on the results of the population detection steps in (b).}
    \label{fig:spam-protocol}
\end{figure*}

Consider a multi-level quantum system with two long-lived manifolds, \textit{A} and \textit{B}, connected by population transfer channels, schematically represented in Fig.~\ref{fig:spam-protocol}(a). The aim of our protocol is to prepare and measure a qubit encoded in states $\{\ket{0}, \ket{1}\}$ somewhere within $A$ and $B$. This problem setting is motivated by the structure of many atomic qubits, where laser pulses can be used to transfer population between ground states (in $A$) and metastable states (in $B$) \cite{schmidt-kaler_how_2003, schmidt-kaler_coherence_2003}. Depending on the application, it may be desirable to encode both qubit states in the ground level (qubit type $G$), or in the metastable level (qubit type $M$), or to place one qubit state in $A$ and one in $B$ (qubit type $O$) \cite{allcock_omg_2021}. In this section, we describe the SPAM protocol in a general manner, independent of the quantum system or qubit encoding. In the subsequent section, we demonstrate its performance with all three qubit encodings in a single trapped \ba ion.

Our protocol relies on three capabilities. First, we must be able to detect the population within $A$ using some physical process that does not couple to states within $B$. Henceforth, we will call this process \textit{population detection} to distinguish it from quantum measurement. We label the outcome of population detection as \textit{bright} if we detect population within $A$ and as \textit{dark} otherwise. Second, we must be able to transfer the population between individual states in $A$ and $B$ (see below). Third, we are ideally able to initialize the majority of the population in some state $\ket{SP}_A$ in $A$. In our protocol, population detection is used to flag errors during population initialization and population transfer, as well as $B \rightarrow A$ leakage errors that leave population outside of the qubit manifold. As a result, the SPAM fidelity depends predominantly on the population detection fidelity, while errors during other steps mainly lead to an effective slowdown of computation.

The full SPAM sequence, implemented in a quantum system with the above properties, is shown in Fig.~\ref{fig:spam-protocol}(b). We split the protocol into a \textit{state preparation} part and a \textit{measurement} part, in between which a desired quantum computation would be implemented. The state preparation part of the protocol starts with initializing the system in a well-defined state $\ket{SP}_A$ using standard methods --- for example, optical pumping in the case of an atomic system. This is followed by transferring the population from $\ket{SP}_A$ to any state in $B$. We then perform a population detection step ($R_1$) and raise an error flag in the case of a bright outcome, signifying population left within $A$. This step thus eliminates errors stemming from both imperfect initialization of the system into $\ket{SP}_A$ and imperfect population transfer from $\ket{SP}_A$ into $B$. Finally, if the state we transferred to in $B$ is not part of the qubit manifold, we apply additional transfer pulses to move the population to either $\ket{0}$ or $\ket{1}$, completing the state preparation sequence.

\begin{figure*}[!ht]
    \includegraphics[width=\textwidth]{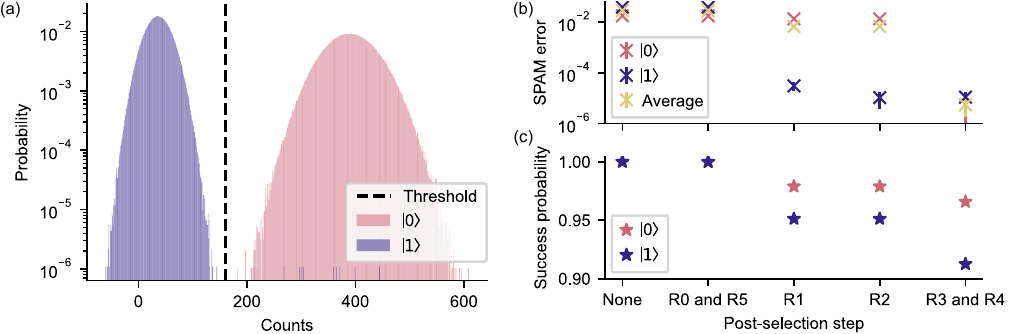}
    \caption{SPAM of the metastable (M) qubit in \ba. The protocol is repeated $10^6$ times for each of the two qubit states. (a) Histogram showing the number of counts collected on the camera during the population detection step $R_3$. Blue (red) bars represent the result for a system initially prepared in $\ket{1}$ ($\ket{0}$). The dashed vertical line shows the threshold for classifying results as either dark or bright. An uncalibrated nominal camera dark count offset of $100$ counts per pixel has been subtracted from each raw count value, leading to occasional negative count readings. (b) SPAM error after each post-selection step. After the final step, the measured SPAM error is $0.0(2.0)\times 10^{-6}$ for $\ket{0}$ and $1.1(7)\times 10^{-5}$ for $\ket{1}$, giving an average error of $5(4)\times 10^{-6}$. The uncertainties represent one Wilson interval. (c) Probability that a shot is kept as a function of the post-selection step. We find that, while $<10\%$ of the shots are rejected, the SPAM error is reduced by four orders of magnitude. The differences between the two states arise from differing fidelities of the various $1762\nm$ pulses used to transfer the qubit states to and from the \ground level. }
    \label{fig:spam-data-metastable}
\end{figure*}

We now describe the measurement part of the protocol. If the qubit is not fully encoded within $B$, we apply transfer pulses to move the population from $\ket{0}$ into $\ket{0}_B$ and from $\ket{1}$ into $\ket{1}_B$, where $\ket{0}_B$ and $\ket{1}_B$ are both in $B$ as in Fig.~\ref{fig:spam-protocol}(a). We then perform a population detection step, $R_2$ in Fig.~\ref{fig:spam-protocol}(b). At this stage, we expect all population to be in $B$, so detecting population in $A$ (bright outcome) indicates an error. This step flags errors due to leakage from the qubit manifold into $A$ -- for example, the spontaneous decay of $B$ into $A$ \cite{schmidt-kaler_how_2003}, as well as errors due to imperfect population transfer in the previous steps. After this step, the population in $\ket{0}_B$ is transferred to a state in $A$ and a population detection step, $R_3$ in Fig.~\ref{fig:spam-protocol}(b), is carried out again. The outcome is used to infer the \textit{quantum measurement result}, where a bright outcome corresponds to the measurement result $\ket{0}$ and a dark outcome corresponds to the measurement result $\ket{1}$. Finally, the $\ket{1}_B$ state is transferred to a state in $A$ and a final population detection step, $R_4$ in Fig.~\ref{fig:spam-protocol}(b), is carried out. As the qubit has to be either in the state $\ket{0}$ or the state $\ket{1}$, either $R_3$ or $R_4$ has to result in a bright measurement outcome. Measuring dark both times suggests that the population is left in $B$ due to imperfect population transfer to $A$, leakage out of the qubit manifold but still within $B$, or a total qubit loss. In any case, the error flag is raised.

The complete truth table for deciding whether to raise a flag based on the outcomes of the four population detection steps is shown in Fig.~\ref{fig:spam-protocol}(c). This protocol allows us to detect the most common sources of error in qubit SPAM. It thus eliminates the dependence of the SPAM fidelity on the population transfer and population initialization quality and makes it dependent only on the population detection quality, which can be made very high and very robust to drifts \cite{burrell_scalable_2010}.

Assuming perfect population detection, there are only two types of physical errors that the protocol cannot detect. The first is population transfer in or out of the wrong qubit state, e.g. mapping $\ket{0}$ onto $\ket{1}_B$ instead of $\ket{0}_B$. The second is the leakage of population from $\ket{1}_B$ to $A$ between $R_2$ and $R_3$ or during $R_3$. This would result in a bright outcome of $R_3$, leading us to wrongly conclude a measurement result $\ket{0}$. 

There are two possible ways of dealing with shots where the error flag was raised. First, it is possible to let the computation continue deterministically and discard the results of flagged shots in the data analysis stage, a technique known as post-selection \cite{aaronson_quantum_2004}. Alternatively, the computation flow can be adapted in real-time, for example, by repeating state preparation until success, or up to a fixed maximum number of attempts \cite{lim_repeat-until-success_2005}. While the measurement sequence cannot be repeated until success, it can be used to detect errors and convert them into erasures, which are some of the most straightforward errors to correct using QEC \cite{gottesman_stabilizer_1997, grassl_codes_1997, bennett_capacities_1997, wu_erasure_2022}. The real-time adaptive control minimizes the computational slowdown caused by data rejection, which makes it particularly useful for deep circuits and large registers. On the other hand, deterministic operation with post-selection poses minimal requirements on the experimental control system and allows for the lowest error rates, making it useful for near-term algorithms with small-scale QCs.

\section{Protocol implementation}
\label{sec:protocol-implementation}
We implement the SPAM protocol in a single \ba\; ion in a Paul trap, using an experimental setup similar to the one previously described in \cite{choonee_silicon_2017, wilpers_compact_2013, sotirova_low_2024} (see also Sec.~\ref{sec:si-exp-setup}). We identify
the manifolds $A$ and $B$ with the levels \ground and \shelf in \ba, respectively (see Sec.~\ref{sec:si-ba-level-structure} for a complete level diagram). We benchmark the protocol on the three aforementioned qubit types, using encodings O: $\ket{D_{5/2}, F=2, m_F=-1}\leftrightarrow\ket{S_{1/2}, F=2, m_F=0}$,
M: $\ket{D_{5/2}, F=2, m_F=-1}\leftrightarrow\ket{D_{5/2}, F=1, m_F=-1}$, and G: $\ket{S_{1/2}, F=2, m_F=0}\leftrightarrow\ket{S_{1/2}, F=1, m_F=0}$.

Population initialization is performed by turning on $\pi$-polarized $493\nm$ light resonant with the $\ket{S_{1/2}, F=1} \leftrightarrow \ket{P_{1/2}, F=2}$ and $\ket{S_{1/2}, F=2} \leftrightarrow \ket{P_{1/2}, F=2}$ transitions for $20\us$ (see Sec.~\ref{sec:si-ba-level-structure}) \cite{an_high_2022}. We measure a state initialization error of $0.8(6)\%$ (see Sec.~\ref{sec:si-data-rate}), which we attribute to polarization impurity of the $493\nm$ beam and off-resonant excitation of the $\ket{S_{1/2}, F=2} \leftrightarrow \ket{P_{1/2}, F=1}$ transition.

Population transfer is performed using coherent laser pulses at $1762\nm$ resonant with individual \ground $\leftrightarrow$ \shelf transitions. The specific transitions addressed differ depending on the qubit encoding and are listed in detail in Sec.~\ref{sec:si-spam-sequence}. The average duration of the $1762\nm$ pulses is $ \approx25\us$. Individual transfer pulses were measured to have an error rate in the range of $1\%-5\%$ (see Sec.~\ref{sec:si-data-rate}), largely limited by magnetic field noise and laser phase noise. Further details on the experimental setup can be found in Sec.~\ref{sec:si-exp-setup} and Ref.~\cite{sotirova_thesis_2024}.

Population detection is performed using a $493\nm$ laser to continuously drive the cycling $\mathrm{S_{1/2}}\leftrightarrow\mathrm{P_{1/2}}$ transition and imaging the scattered photons on a scientific CMOS (sCMOS) camera (see Sec.~\ref{sec:si-exp-setup}). We use a $493\nm$ laser detuning of $\approx -5\MHz$ relative to the measured line centre \cite{leupold_sustained_2018} and a power of $\approx 100I_\mathrm{sat}$, where the saturation intensity $I_\mathrm{sat}$ is as defined in Ref.~\cite{szwer_high_2009} (see Sec.~\ref{sec:si-optical-detection} and Ref.~\cite{sotirova_thesis_2024}). We collect the photons emitted by the ion on a camera. The camera exposure time is $400\us$. We categorize the outcome of a population detection as bright or dark by comparing the number of camera counts to a pre-calibrated threshold (see Sec.~\ref{sec:si-optical-detection} and Ref.~\cite{sotirova_thesis_2024}). There are two contributions to the population detection error. The first is the optical detection fidelity -- the probability that the detection of an ion in the \ground level results in a bright outcome, and the probability that the detection of an ion in the \shelf level, which remained in the \shelf level during the detection pulse (or equivalently the lack of an ion), results in a dark outcome. We measure this as described in Sec.~\ref{sec:si-error-budget}, finding an optical detection error of $3.2(28)\times 10^{-6}$ on the bright state and $0.0(1.1)\times10^{-6}$ on the dark state. The second contribution is the probability of decay of the metastable \shelf level during the population detection pulse, which we measure as $1.4(6)\times10^{-5}$. This agrees with the measured lifetime of the metastable level of $27.2(2)\s$ (see Sec.~\ref{sec:si-metastable-decay}). The above measurements lead to an expected average population detection error of $8.6(34)\times10^{-6}$ (see Sec.~\ref{sec:si-error-budget} for details).

We benchmark the SPAM protocol using a sequence similar to Fig.~\ref{fig:spam-protocol}, with a few modifications. At the start of each shot, the ion is Doppler-cooled \cite{neuhauser_optical-sideband_1978, wineland_radiation-pressure_1978, wineland_laser_1979}, and a population detection step $R_0$ is executed to verify the presence of an ion. We then Doppler-cool the ion again and perform the state preparation sequence, followed by the measurement sequence. At the end of the protocol, we deshelve the population out of \shelf (see Sec.~\ref{sec:si-ba-level-structure}) and perform a final population detection step $R_5$ to confirm the ion remained in the trap during the experiment. Hence, in addition to the rules in the table in Fig.~\ref{fig:spam-protocol}(c), we also flag any shots where the result of either $R_0$ or $R_5$ is dark. Afterwards, we discard the shots where a flag was raised using post-selection. We interleave the attempts to prepare $\ket{0}$ with attempts to prepare $\ket{1}$ and record the SPAM fidelity, which is the probability that the measurement outcome matches the intended value \cite{burrell_high_2010}. Despite performing a total of six population detection steps per shot, we expect the average SPAM error to be equal to that of one population detection, i.e. $8.6(34)\times10^{-6}$. The detailed error budget for the protocol is presented in Sec.~\ref{sec:si-error-budget}.

\begin{figure}
    \centering
    \includegraphics[width=1\linewidth]{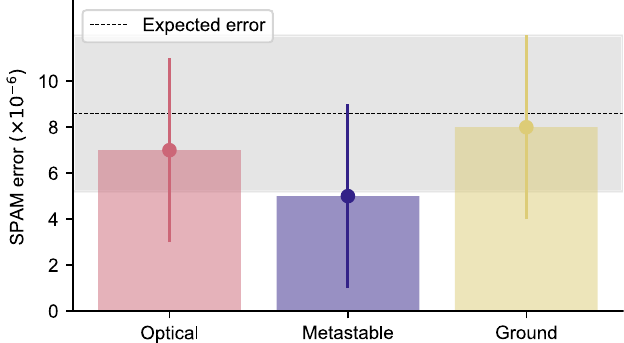}
    \caption{Measured SPAM errors for all three qubit types in \ba --- O: $7(4) \times 10^{-6}$, M: $5(4) \times 10^{-6}$, and G: $8(4) \times 10^{-6}$. The errors for all three qubit types agree within the error bars. The measured errors also agree with the expected error given independent measurements of the individual error sources (see Section \ref{sec:si-error-budget} for further details).}
    \label{fig:spam-all-qubits}
\end{figure}

The experimental results for the metastable qubit in \ba are summarised in Fig.~\ref{fig:spam-data-metastable}. We repeat the protocol $10^6$ times for each of the two qubit states. In Fig.~\ref{fig:spam-data-metastable}(a), we show the observed counts during population detection $R_3$ when preparing $\ket{0}$ and $\ket{1}$ after post-selection according to the truth table in Fig.~\ref{fig:spam-protocol}(c). The histograms demonstrate that we can achieve excellent population detection in our experiment, which is the core requirement for the high-fidelity implementation of our protocol. In Fig.~\ref{fig:spam-data-metastable}(b) we show the reduction of the average SPAM error after each post-selection step. We find that, without any post-selection, this data exhibits a SPAM error of $3.86(4)\times 10^{-2}$ for $\ket{1}$ and $1.754(26)\times 10^{-2}$ for $\ket{0}$, giving an average error of $4.126(27)\times 10^{-2}$. After all post-selection steps, the SPAM error is reduced down to the final value of $1.1(7)\times10^{-5}$ for $\ket{1}$ and $0.0(20)\times10^{-6}$ for $\ket{0}$, giving an average SPAM error of $5(4)\times10^{-6}$. This agrees with the expected error of $8.6(34)\times 10^{-6}$. The measured error asymmetry for $\ket{0}$ and $\ket{1}$ is caused by the fact that \shelf can decay to \ground but not the other way around. This asymmetry can be exploited to reduce average SPAM errors in QC applications where different measurement outcomes are not equally likely, such as in QEC \cite{leung_approximate_1997, kubica_erasure_2022}.

In Fig.~\ref{fig:spam-data-metastable}(c), we plot the SPAM success probability (or equivalently the effective computation slowdown rate) after each post-selection step. We find that, while post-section reduces the SPAM error by over four orders of magnitude, it does so while discarding $< 10\%$ of experimental shots. We note that, due to the different $1762\nm$ population transfer pulses experiencing different errors (see Sec.~\ref{sec:si-data-rate}), qubit states are discarded with unequal probabilities. This does not affect the SPAM fidelity but can bias the results of quantum computing applications that require estimating observable averages \cite{itano_quantum_1993, cao_potential_2018, cerezo_variational_2021, ferguson_measurement-based_2021}. This issue can be corrected by alternating the definition of $\ket{0}$ and $\ket{1}$ for each shot or by carefully measuring the data rejection rate for each qubit state and inverting it in post-processing, see Sec.~\ref{sec:superposition-readout} for details.

Finally, we repeat the SPAM error measurement for the ground and optical qubit encodings. The results are shown in Fig.~\ref{fig:spam-all-qubits}. As expected from the model, we find that the SPAM fidelities on all three qubit types are the same within experimental uncertainty. Averaged over all three encodings, we record a mean SPAM error of $7(3) \times 10^{-6}$, a record for any qubit in any platform. This demonstrates the encoding flexibility offered by our protocol -- as long as a metastable manifold is available, it can be exploited for high-fidelity SPAM, regardless of whether it is part of the qubit encoding.

\section{Conclusion}
\label{sec:conclusion}
We have developed a new protocol for qubit SPAM that uses a series of mid-circuit QND measurements to effectively detect and flag errors. As a result, the SPAM fidelity depends predominantly on the QND measurement fidelity, which can typically be made very high and robust. Our experimental results with a single trapped \ba ion show that this protocol achieves state-of-the-art performance across three different qubit encodings, demonstrating its versatility.

The only requirement for applying this protocol is a quantum system with a metastable structure as shown Fig.~\ref{fig:spam-protocol}(a), which is common to many quantum computing platforms. This makes the protocol highly general, and applicable to a wide range of systems -- including both natural and artificial atoms \cite{astafiev_resonance_2010, cottet_electron_2021} -- and encodings, whether based on qubits or qudits \cite{low_practical_2020, ringbauer_universal_2022, low_control_2023, hrmo_native_2023, goss_extending_2023}. Our protocol can also be applied to atomic and molecular experiments beyond QC, e.g. in precision measurements \cite{leibfried_quantum_2012, wolf_prospect_2024}. Finally, as the capabilities needed to implement the individual steps of our protocol are typically already available in QC setups, we anticipate that our method can be straightforwardly deployed to improve the performance of many existing systems.

\setcounter{secnumdepth}{0}

\section{Acknowledgments}
The authors thank Peter Drmota, David Nadlinger, David Allcock, and David Lucas for helpful discussions. This work was supported by a UKRI FL Fellowship (MR/S03238X/1). ASS acknowledges funding from the JT Hamilton scholarship from Balliol College, Oxford. SMD acknowledges funding from the Jowett scholarship from Balliol College, Oxford.

\section{Competing interests}
ASS, FP, and MM are employees of Oxford Ionics Ltd. CJB is a director of Oxford Ionics Ltd. The remaining authors declare no competing interests.

\bibliographystyle{apsrev4-1}
\bibliography{references}

\begin{thebibliography}{58}%
\makeatletter
\providecommand \@ifxundefined [1]{%
 \@ifx{#1\undefined}
}%
\providecommand \@ifnum [1]{%
 \ifnum #1\expandafter \@firstoftwo
 \else \expandafter \@secondoftwo
 \fi
}%
\providecommand \@ifx [1]{%
 \ifx #1\expandafter \@firstoftwo
 \else \expandafter \@secondoftwo
 \fi
}%
\providecommand \natexlab [1]{#1}%
\providecommand \enquote  [1]{``#1''}%
\providecommand \bibnamefont  [1]{#1}%
\providecommand \bibfnamefont [1]{#1}%
\providecommand \citenamefont [1]{#1}%
\providecommand \href@noop [0]{\@secondoftwo}%
\providecommand \href [0]{\begingroup \@sanitize@url \@href}%
\providecommand \@href[1]{\@@startlink{#1}\@@href}%
\providecommand \@@href[1]{\endgroup#1\@@endlink}%
\providecommand \@sanitize@url [0]{\catcode `\\12\catcode `\$12\catcode `\&12\catcode `\#12\catcode `\^12\catcode `\_12\catcode `\%12\relax}%
\providecommand \@@startlink[1]{}%
\providecommand \@@endlink[0]{}%
\providecommand \url  [0]{\begingroup\@sanitize@url \@url }%
\providecommand \@url [1]{\endgroup\@href {#1}{\urlprefix }}%
\providecommand \urlprefix  [0]{URL }%
\providecommand \Eprint [0]{\href }%
\providecommand \doibase [0]{http://dx.doi.org/}%
\providecommand \selectlanguage [0]{\@gobble}%
\providecommand \bibinfo  [0]{\@secondoftwo}%
\providecommand \bibfield  [0]{\@secondoftwo}%
\providecommand \translation [1]{[#1]}%
\providecommand \BibitemOpen [0]{}%
\providecommand \bibitemStop [0]{}%
\providecommand \bibitemNoStop [0]{.\EOS\space}%
\providecommand \EOS [0]{\spacefactor3000\relax}%
\providecommand \BibitemShut  [1]{\csname bibitem#1\endcsname}%
\let\auto@bib@innerbib\@empty
\bibitem [{\citenamefont {DiVincenzo}(2000)}]{divincenzo_physical_2000}%
  \BibitemOpen
  \bibfield  {author} {\bibinfo {author} {\bibfnamefont {D.~P.}\ \bibnamefont {DiVincenzo}},\ }\href {\doibase 10.1002/1521-3978(200009)48:9/11<771::AID-PROP771>3.0.CO;2-E} {\bibfield  {journal} {\bibinfo  {journal} {Fortschritte der Physik}\ }\textbf {\bibinfo {volume} {48}},\ \bibinfo {pages} {771} (\bibinfo {year} {2000})}\BibitemShut {NoStop}%
\bibitem [{\citenamefont {Kastler}(1950)}]{kastler_quelques_1950}%
  \BibitemOpen
  \bibfield  {author} {\bibinfo {author} {\bibfnamefont {A.}~\bibnamefont {Kastler}},\ }\href {\doibase 10.1051/jphysrad:01950001106025500} {\bibfield  {journal} {\bibinfo  {journal} {Journal de Physique et le Radium}\ }\textbf {\bibinfo {volume} {11}},\ \bibinfo {pages} {255} (\bibinfo {year} {1950})}\BibitemShut {NoStop}%
\bibitem [{\citenamefont {Brossel}\ \emph {et~al.}(1952)\citenamefont {Brossel}, \citenamefont {Kastler},\ and\ \citenamefont {Winter}}]{brossel_greation_1952}%
  \BibitemOpen
  \bibfield  {author} {\bibinfo {author} {\bibfnamefont {J.}~\bibnamefont {Brossel}}, \bibinfo {author} {\bibfnamefont {A.}~\bibnamefont {Kastler}}, \ and\ \bibinfo {author} {\bibfnamefont {J.}~\bibnamefont {Winter}},\ }\href {\doibase 10.1051/jphysrad:019520013012066800} {\bibfield  {journal} {\bibinfo  {journal} {Journal de Physique et le Radium}\ }\textbf {\bibinfo {volume} {13}},\ \bibinfo {pages} {668} (\bibinfo {year} {1952})}\BibitemShut {NoStop}%
\bibitem [{\citenamefont {Hawkins}\ and\ \citenamefont {Dicke}(1953)}]{hawkins_polarization_1953}%
  \BibitemOpen
  \bibfield  {author} {\bibinfo {author} {\bibfnamefont {W.~B.}\ \bibnamefont {Hawkins}}\ and\ \bibinfo {author} {\bibfnamefont {R.~H.}\ \bibnamefont {Dicke}},\ }\href {\doibase 10.1103/PhysRev.91.1008} {\bibfield  {journal} {\bibinfo  {journal} {Physical Review}\ }\textbf {\bibinfo {volume} {91}},\ \bibinfo {pages} {1008} (\bibinfo {year} {1953})}\BibitemShut {NoStop}%
\bibitem [{\citenamefont {Acton}\ \emph {et~al.}(2006)\citenamefont {Acton}, \citenamefont {Brickman}, \citenamefont {Haljan}, \citenamefont {Lee}, \citenamefont {Deslauriers},\ and\ \citenamefont {Monroe}}]{acton_near-perfect_2006}%
  \BibitemOpen
  \bibfield  {author} {\bibinfo {author} {\bibfnamefont {M.}~\bibnamefont {Acton}}, \bibinfo {author} {\bibfnamefont {K.-A.}\ \bibnamefont {Brickman}}, \bibinfo {author} {\bibfnamefont {P.~C.}\ \bibnamefont {Haljan}}, \bibinfo {author} {\bibfnamefont {P.~J.}\ \bibnamefont {Lee}}, \bibinfo {author} {\bibfnamefont {L.}~\bibnamefont {Deslauriers}}, \ and\ \bibinfo {author} {\bibfnamefont {C.}~\bibnamefont {Monroe}},\ }\href {http://arxiv.org/abs/quant-ph/0511257} {\enquote {\bibinfo {title} {Near-{Perfect} {Simultaneous} {Measurement} of a {Qubit} {Register}},}\ } (\bibinfo {year} {2006}),\ \bibinfo {note} {arXiv:quant-ph/0511257}\BibitemShut {NoStop}%
\bibitem [{\citenamefont {Nagourney}\ \emph {et~al.}(1986)\citenamefont {Nagourney}, \citenamefont {Sandberg},\ and\ \citenamefont {Dehmelt}}]{nagourney_shelved_1986}%
  \BibitemOpen
  \bibfield  {author} {\bibinfo {author} {\bibfnamefont {W.}~\bibnamefont {Nagourney}}, \bibinfo {author} {\bibfnamefont {J.}~\bibnamefont {Sandberg}}, \ and\ \bibinfo {author} {\bibfnamefont {H.}~\bibnamefont {Dehmelt}},\ }\href {\doibase 10.1103/PhysRevLett.56.2797} {\bibfield  {journal} {\bibinfo  {journal} {Physical Review Letters}\ }\textbf {\bibinfo {volume} {56}},\ \bibinfo {pages} {2797} (\bibinfo {year} {1986})}\BibitemShut {NoStop}%
\bibitem [{\citenamefont {Sauter}\ \emph {et~al.}(1986)\citenamefont {Sauter}, \citenamefont {Neuhauser}, \citenamefont {Blatt},\ and\ \citenamefont {Toschek}}]{sauter_observation_1986}%
  \BibitemOpen
  \bibfield  {author} {\bibinfo {author} {\bibfnamefont {T.}~\bibnamefont {Sauter}}, \bibinfo {author} {\bibfnamefont {W.}~\bibnamefont {Neuhauser}}, \bibinfo {author} {\bibfnamefont {R.}~\bibnamefont {Blatt}}, \ and\ \bibinfo {author} {\bibfnamefont {P.~E.}\ \bibnamefont {Toschek}},\ }\href {\doibase 10.1103/PhysRevLett.57.1696} {\bibfield  {journal} {\bibinfo  {journal} {Physical Review Letters}\ }\textbf {\bibinfo {volume} {57}},\ \bibinfo {pages} {1696} (\bibinfo {year} {1986})}\BibitemShut {NoStop}%
\bibitem [{\citenamefont {Bergquist}\ \emph {et~al.}(1986)\citenamefont {Bergquist}, \citenamefont {Hulet}, \citenamefont {Itano},\ and\ \citenamefont {Wineland}}]{bergquist_observation_1986}%
  \BibitemOpen
  \bibfield  {author} {\bibinfo {author} {\bibfnamefont {J.~C.}\ \bibnamefont {Bergquist}}, \bibinfo {author} {\bibfnamefont {R.~G.}\ \bibnamefont {Hulet}}, \bibinfo {author} {\bibfnamefont {W.~M.}\ \bibnamefont {Itano}}, \ and\ \bibinfo {author} {\bibfnamefont {D.~J.}\ \bibnamefont {Wineland}},\ }\href {\doibase 10.1103/PhysRevLett.57.1699} {\bibfield  {journal} {\bibinfo  {journal} {Physical Review Letters}\ }\textbf {\bibinfo {volume} {57}},\ \bibinfo {pages} {1699} (\bibinfo {year} {1986})}\BibitemShut {NoStop}%
\bibitem [{\citenamefont {Purcell}\ \emph {et~al.}(1946)\citenamefont {Purcell}, \citenamefont {Torrey},\ and\ \citenamefont {Pound}}]{purcell_resonance_1946}%
  \BibitemOpen
  \bibfield  {author} {\bibinfo {author} {\bibfnamefont {E.~M.}\ \bibnamefont {Purcell}}, \bibinfo {author} {\bibfnamefont {H.~C.}\ \bibnamefont {Torrey}}, \ and\ \bibinfo {author} {\bibfnamefont {R.~V.}\ \bibnamefont {Pound}},\ }\href {\doibase 10.1103/PhysRev.69.37} {\bibfield  {journal} {\bibinfo  {journal} {Physical Review}\ }\textbf {\bibinfo {volume} {69}},\ \bibinfo {pages} {37} (\bibinfo {year} {1946})}\BibitemShut {NoStop}%
\bibitem [{\citenamefont {Reed}\ \emph {et~al.}(2010)\citenamefont {Reed}, \citenamefont {Johnson}, \citenamefont {Houck}, \citenamefont {DiCarlo}, \citenamefont {Chow}, \citenamefont {Schuster}, \citenamefont {Frunzio},\ and\ \citenamefont {Schoelkopf}}]{reed_fast_2010}%
  \BibitemOpen
  \bibfield  {author} {\bibinfo {author} {\bibfnamefont {M.~D.}\ \bibnamefont {Reed}}, \bibinfo {author} {\bibfnamefont {B.~R.}\ \bibnamefont {Johnson}}, \bibinfo {author} {\bibfnamefont {A.~A.}\ \bibnamefont {Houck}}, \bibinfo {author} {\bibfnamefont {L.}~\bibnamefont {DiCarlo}}, \bibinfo {author} {\bibfnamefont {J.~M.}\ \bibnamefont {Chow}}, \bibinfo {author} {\bibfnamefont {D.~I.}\ \bibnamefont {Schuster}}, \bibinfo {author} {\bibfnamefont {L.}~\bibnamefont {Frunzio}}, \ and\ \bibinfo {author} {\bibfnamefont {R.~J.}\ \bibnamefont {Schoelkopf}},\ }\href {\doibase 10.1063/1.3435463} {\bibfield  {journal} {\bibinfo  {journal} {Applied Physics Letters}\ }\textbf {\bibinfo {volume} {96}},\ \bibinfo {pages} {203110} (\bibinfo {year} {2010})}\BibitemShut {NoStop}%
\bibitem [{\citenamefont {Walter}\ \emph {et~al.}(2017)\citenamefont {Walter}, \citenamefont {Kurpiers}, \citenamefont {Gasparinetti}, \citenamefont {Magnard}, \citenamefont {Potočnik}, \citenamefont {Salathé}, \citenamefont {Pechal}, \citenamefont {Mondal}, \citenamefont {Oppliger}, \citenamefont {Eichler},\ and\ \citenamefont {Wallraff}}]{walter_rapid_2017}%
  \BibitemOpen
  \bibfield  {author} {\bibinfo {author} {\bibfnamefont {T.}~\bibnamefont {Walter}}, \bibinfo {author} {\bibfnamefont {P.}~\bibnamefont {Kurpiers}}, \bibinfo {author} {\bibfnamefont {S.}~\bibnamefont {Gasparinetti}}, \bibinfo {author} {\bibfnamefont {P.}~\bibnamefont {Magnard}}, \bibinfo {author} {\bibfnamefont {A.}~\bibnamefont {Potočnik}}, \bibinfo {author} {\bibfnamefont {Y.}~\bibnamefont {Salathé}}, \bibinfo {author} {\bibfnamefont {M.}~\bibnamefont {Pechal}}, \bibinfo {author} {\bibfnamefont {M.}~\bibnamefont {Mondal}}, \bibinfo {author} {\bibfnamefont {M.}~\bibnamefont {Oppliger}}, \bibinfo {author} {\bibfnamefont {C.}~\bibnamefont {Eichler}}, \ and\ \bibinfo {author} {\bibfnamefont {A.}~\bibnamefont {Wallraff}},\ }\href {\doibase 10.1103/PhysRevApplied.7.054020} {\bibfield  {journal} {\bibinfo  {journal} {Physical Review Applied}\ }\textbf {\bibinfo {volume} {7}},\ \bibinfo {pages} {054020} (\bibinfo {year} {2017})}\BibitemShut {NoStop}%
\bibitem [{\citenamefont {Myerson}\ \emph {et~al.}(2008)\citenamefont {Myerson}, \citenamefont {Szwer}, \citenamefont {Webster}, \citenamefont {Allcock}, \citenamefont {Curtis}, \citenamefont {Imreh}, \citenamefont {Sherman}, \citenamefont {Stacey}, \citenamefont {Steane},\ and\ \citenamefont {Lucas}}]{myerson_high-fidelity_2008}%
  \BibitemOpen
  \bibfield  {author} {\bibinfo {author} {\bibfnamefont {A.~H.}\ \bibnamefont {Myerson}}, \bibinfo {author} {\bibfnamefont {D.~J.}\ \bibnamefont {Szwer}}, \bibinfo {author} {\bibfnamefont {S.~C.}\ \bibnamefont {Webster}}, \bibinfo {author} {\bibfnamefont {D.~T.~C.}\ \bibnamefont {Allcock}}, \bibinfo {author} {\bibfnamefont {M.~J.}\ \bibnamefont {Curtis}}, \bibinfo {author} {\bibfnamefont {G.}~\bibnamefont {Imreh}}, \bibinfo {author} {\bibfnamefont {J.~A.}\ \bibnamefont {Sherman}}, \bibinfo {author} {\bibfnamefont {D.~N.}\ \bibnamefont {Stacey}}, \bibinfo {author} {\bibfnamefont {A.~M.}\ \bibnamefont {Steane}}, \ and\ \bibinfo {author} {\bibfnamefont {D.~M.}\ \bibnamefont {Lucas}},\ }\href {\doibase 10.1103/PhysRevLett.100.200502} {\bibfield  {journal} {\bibinfo  {journal} {Physical Review Letters}\ }\textbf {\bibinfo {volume} {100}},\ \bibinfo {pages} {200502} (\bibinfo {year} {2008})}\BibitemShut {NoStop}%
\bibitem [{\citenamefont {An}\ \emph {et~al.}(2022)\citenamefont {An}, \citenamefont {Ransford}, \citenamefont {Schaffer}, \citenamefont {Sletten}, \citenamefont {Gaebler}, \citenamefont {Hostetter},\ and\ \citenamefont {Vittorini}}]{an_high_2022}%
  \BibitemOpen
  \bibfield  {author} {\bibinfo {author} {\bibfnamefont {F.~A.}\ \bibnamefont {An}}, \bibinfo {author} {\bibfnamefont {A.}~\bibnamefont {Ransford}}, \bibinfo {author} {\bibfnamefont {A.}~\bibnamefont {Schaffer}}, \bibinfo {author} {\bibfnamefont {L.~R.}\ \bibnamefont {Sletten}}, \bibinfo {author} {\bibfnamefont {J.}~\bibnamefont {Gaebler}}, \bibinfo {author} {\bibfnamefont {J.}~\bibnamefont {Hostetter}}, \ and\ \bibinfo {author} {\bibfnamefont {G.}~\bibnamefont {Vittorini}},\ }\href {\doibase 10.1103/PhysRevLett.129.130501} {\bibfield  {journal} {\bibinfo  {journal} {Physical Review Letters}\ }\textbf {\bibinfo {volume} {129}},\ \bibinfo {pages} {130501} (\bibinfo {year} {2022})}\BibitemShut {NoStop}%
\bibitem [{\citenamefont {Harty}\ \emph {et~al.}(2014)\citenamefont {Harty}, \citenamefont {Allcock}, \citenamefont {Ballance}, \citenamefont {Guidoni}, \citenamefont {Janacek}, \citenamefont {Linke}, \citenamefont {Stacey},\ and\ \citenamefont {Lucas}}]{harty_high-fidelity_2014}%
  \BibitemOpen
  \bibfield  {author} {\bibinfo {author} {\bibfnamefont {T.}~\bibnamefont {Harty}}, \bibinfo {author} {\bibfnamefont {D.}~\bibnamefont {Allcock}}, \bibinfo {author} {\bibfnamefont {C.}~\bibnamefont {Ballance}}, \bibinfo {author} {\bibfnamefont {L.}~\bibnamefont {Guidoni}}, \bibinfo {author} {\bibfnamefont {H.}~\bibnamefont {Janacek}}, \bibinfo {author} {\bibfnamefont {N.}~\bibnamefont {Linke}}, \bibinfo {author} {\bibfnamefont {D.}~\bibnamefont {Stacey}}, \ and\ \bibinfo {author} {\bibfnamefont {D.}~\bibnamefont {Lucas}},\ }\href {\doibase 10.1103/PhysRevLett.113.220501} {\bibfield  {journal} {\bibinfo  {journal} {Physical Review Letters}\ }\textbf {\bibinfo {volume} {113}},\ \bibinfo {pages} {220501} (\bibinfo {year} {2014})}\BibitemShut {NoStop}%
\bibitem [{\citenamefont {Leu}\ \emph {et~al.}(2024)\citenamefont {Leu}, \citenamefont {Smith}, \citenamefont {Gely},\ and\ \citenamefont {Lucas}}]{leu_polarisation-insensitive_2024}%
  \BibitemOpen
  \bibfield  {author} {\bibinfo {author} {\bibfnamefont {A.~D.}\ \bibnamefont {Leu}}, \bibinfo {author} {\bibfnamefont {M.~C.}\ \bibnamefont {Smith}}, \bibinfo {author} {\bibfnamefont {M.~F.}\ \bibnamefont {Gely}}, \ and\ \bibinfo {author} {\bibfnamefont {D.~M.}\ \bibnamefont {Lucas}},\ }\href {http://arxiv.org/abs/2406.14448} {\enquote {\bibinfo {title} {Polarisation-insensitive state preparation for trapped-ion hyperfine qubits},}\ } (\bibinfo {year} {2024}),\ \bibinfo {note} {arXiv:2406.14448}\BibitemShut {NoStop}%
\bibitem [{\citenamefont {Guggemos}\ \emph {et~al.}(2019)\citenamefont {Guggemos}, \citenamefont {Guevara-Bertsch}, \citenamefont {Heinrich}, \citenamefont {Herrera-Sancho}, \citenamefont {Colombe}, \citenamefont {Blatt},\ and\ \citenamefont {Roos}}]{guggemos_frequency_2019}%
  \BibitemOpen
  \bibfield  {author} {\bibinfo {author} {\bibfnamefont {M.}~\bibnamefont {Guggemos}}, \bibinfo {author} {\bibfnamefont {M.}~\bibnamefont {Guevara-Bertsch}}, \bibinfo {author} {\bibfnamefont {D.}~\bibnamefont {Heinrich}}, \bibinfo {author} {\bibfnamefont {O.~A.}\ \bibnamefont {Herrera-Sancho}}, \bibinfo {author} {\bibfnamefont {Y.}~\bibnamefont {Colombe}}, \bibinfo {author} {\bibfnamefont {R.}~\bibnamefont {Blatt}}, \ and\ \bibinfo {author} {\bibfnamefont {C.~F.}\ \bibnamefont {Roos}},\ }\href {\doibase 10.1088/1367-2630/ab447a} {\bibfield  {journal} {\bibinfo  {journal} {New Journal of Physics}\ }\textbf {\bibinfo {volume} {21}},\ \bibinfo {pages} {103003} (\bibinfo {year} {2019})}\BibitemShut {NoStop}%
\bibitem [{\citenamefont {Grangier}\ \emph {et~al.}(1998)\citenamefont {Grangier}, \citenamefont {Levenson},\ and\ \citenamefont {Poizat}}]{grangier_quantum_1998}%
  \BibitemOpen
  \bibfield  {author} {\bibinfo {author} {\bibfnamefont {P.}~\bibnamefont {Grangier}}, \bibinfo {author} {\bibfnamefont {J.~A.}\ \bibnamefont {Levenson}}, \ and\ \bibinfo {author} {\bibfnamefont {J.-P.}\ \bibnamefont {Poizat}},\ }\href {\doibase 10.1038/25059} {\bibfield  {journal} {\bibinfo  {journal} {Nature}\ }\textbf {\bibinfo {volume} {396}},\ \bibinfo {pages} {537} (\bibinfo {year} {1998})}\BibitemShut {NoStop}%
\bibitem [{\citenamefont {Negnevitsky}\ \emph {et~al.}(2018)\citenamefont {Negnevitsky}, \citenamefont {Marinelli}, \citenamefont {Mehta}, \citenamefont {Lo}, \citenamefont {Flühmann},\ and\ \citenamefont {Home}}]{negnevitsky_repeated_2018}%
  \BibitemOpen
  \bibfield  {author} {\bibinfo {author} {\bibfnamefont {V.}~\bibnamefont {Negnevitsky}}, \bibinfo {author} {\bibfnamefont {M.}~\bibnamefont {Marinelli}}, \bibinfo {author} {\bibfnamefont {K.~K.}\ \bibnamefont {Mehta}}, \bibinfo {author} {\bibfnamefont {H.-Y.}\ \bibnamefont {Lo}}, \bibinfo {author} {\bibfnamefont {C.}~\bibnamefont {Flühmann}}, \ and\ \bibinfo {author} {\bibfnamefont {J.~P.}\ \bibnamefont {Home}},\ }\href {\doibase 10.1038/s41586-018-0668-z} {\bibfield  {journal} {\bibinfo  {journal} {Nature}\ }\textbf {\bibinfo {volume} {563}},\ \bibinfo {pages} {527} (\bibinfo {year} {2018})}\BibitemShut {NoStop}%
\bibitem [{\citenamefont {Aaronson}(2004)}]{aaronson_quantum_2004}%
  \BibitemOpen
  \bibfield  {author} {\bibinfo {author} {\bibfnamefont {S.}~\bibnamefont {Aaronson}},\ }\href {http://arxiv.org/abs/quant-ph/0412187} {\enquote {\bibinfo {title} {Quantum {Computing}, {Postselection}, and {Probabilistic} {Polynomial}-{Time}},}\ } (\bibinfo {year} {2004}),\ \bibinfo {note} {arXiv:quant-ph/0412187}\BibitemShut {NoStop}%
\bibitem [{\citenamefont {Lim}\ \emph {et~al.}(2005)\citenamefont {Lim}, \citenamefont {Beige},\ and\ \citenamefont {Kwek}}]{lim_repeat-until-success_2005}%
  \BibitemOpen
  \bibfield  {author} {\bibinfo {author} {\bibfnamefont {Y.~L.}\ \bibnamefont {Lim}}, \bibinfo {author} {\bibfnamefont {A.}~\bibnamefont {Beige}}, \ and\ \bibinfo {author} {\bibfnamefont {L.~C.}\ \bibnamefont {Kwek}},\ }\href {\doibase 10.1103/PhysRevLett.95.030505} {\bibfield  {journal} {\bibinfo  {journal} {Physical Review Letters}\ }\textbf {\bibinfo {volume} {95}},\ \bibinfo {pages} {030505} (\bibinfo {year} {2005})}\BibitemShut {NoStop}%
\bibitem [{\citenamefont {Nation}\ \emph {et~al.}(2021)\citenamefont {Nation}, \citenamefont {Kang}, \citenamefont {Sundaresan},\ and\ \citenamefont {Gambetta}}]{nation_scalable_2021}%
  \BibitemOpen
  \bibfield  {author} {\bibinfo {author} {\bibfnamefont {P.~D.}\ \bibnamefont {Nation}}, \bibinfo {author} {\bibfnamefont {H.}~\bibnamefont {Kang}}, \bibinfo {author} {\bibfnamefont {N.}~\bibnamefont {Sundaresan}}, \ and\ \bibinfo {author} {\bibfnamefont {J.~M.}\ \bibnamefont {Gambetta}},\ }\href {\doibase 10.1103/PRXQuantum.2.040326} {\bibfield  {journal} {\bibinfo  {journal} {PRX Quantum}\ }\textbf {\bibinfo {volume} {2}},\ \bibinfo {pages} {040326} (\bibinfo {year} {2021})}\BibitemShut {NoStop}%
\bibitem [{\citenamefont {Allcock}\ \emph {et~al.}(2021)\citenamefont {Allcock}, \citenamefont {Campbell}, \citenamefont {Chiaverini}, \citenamefont {Chuang}, \citenamefont {Hudson}, \citenamefont {Moore}, \citenamefont {Ransford}, \citenamefont {Roman}, \citenamefont {Sage},\ and\ \citenamefont {Wineland}}]{allcock_omg_2021}%
  \BibitemOpen
  \bibfield  {author} {\bibinfo {author} {\bibfnamefont {D.~T.~C.}\ \bibnamefont {Allcock}}, \bibinfo {author} {\bibfnamefont {W.~C.}\ \bibnamefont {Campbell}}, \bibinfo {author} {\bibfnamefont {J.}~\bibnamefont {Chiaverini}}, \bibinfo {author} {\bibfnamefont {I.~L.}\ \bibnamefont {Chuang}}, \bibinfo {author} {\bibfnamefont {E.~R.}\ \bibnamefont {Hudson}}, \bibinfo {author} {\bibfnamefont {I.~D.}\ \bibnamefont {Moore}}, \bibinfo {author} {\bibfnamefont {A.}~\bibnamefont {Ransford}}, \bibinfo {author} {\bibfnamefont {C.}~\bibnamefont {Roman}}, \bibinfo {author} {\bibfnamefont {J.~M.}\ \bibnamefont {Sage}}, \ and\ \bibinfo {author} {\bibfnamefont {D.~J.}\ \bibnamefont {Wineland}},\ }\href {\doibase 10.1063/5.0069544} {\bibfield  {journal} {\bibinfo  {journal} {Applied Physics Letters}\ }\textbf {\bibinfo {volume} {119}},\ \bibinfo {pages} {214002} (\bibinfo {year} {2021})}\BibitemShut {NoStop}%
\bibitem [{\citenamefont {Wu}\ \emph {et~al.}(2022)\citenamefont {Wu}, \citenamefont {Kolkowitz}, \citenamefont {Puri},\ and\ \citenamefont {Thompson}}]{wu_erasure_2022}%
  \BibitemOpen
  \bibfield  {author} {\bibinfo {author} {\bibfnamefont {Y.}~\bibnamefont {Wu}}, \bibinfo {author} {\bibfnamefont {S.}~\bibnamefont {Kolkowitz}}, \bibinfo {author} {\bibfnamefont {S.}~\bibnamefont {Puri}}, \ and\ \bibinfo {author} {\bibfnamefont {J.~D.}\ \bibnamefont {Thompson}},\ }\href {\doibase 10.1038/s41467-022-32094-6} {\bibfield  {journal} {\bibinfo  {journal} {Nature Communications}\ }\textbf {\bibinfo {volume} {13}},\ \bibinfo {pages} {4657} (\bibinfo {year} {2022})}\BibitemShut {NoStop}%
\bibitem [{\citenamefont {Kubica}\ \emph {et~al.}(2022)\citenamefont {Kubica}, \citenamefont {Haim}, \citenamefont {Vaknin}, \citenamefont {Brandão},\ and\ \citenamefont {Retzker}}]{kubica_erasure_2022}%
  \BibitemOpen
  \bibfield  {author} {\bibinfo {author} {\bibfnamefont {A.}~\bibnamefont {Kubica}}, \bibinfo {author} {\bibfnamefont {A.}~\bibnamefont {Haim}}, \bibinfo {author} {\bibfnamefont {Y.}~\bibnamefont {Vaknin}}, \bibinfo {author} {\bibfnamefont {F.}~\bibnamefont {Brandão}}, \ and\ \bibinfo {author} {\bibfnamefont {A.}~\bibnamefont {Retzker}},\ }\href {\doibase 10.48550/arXiv.2208.05461} {\enquote {\bibinfo {title} {Erasure qubits: {Overcoming} the ${T}_1$ limit in superconducting circuits},}\ } (\bibinfo {year} {2022}),\ \bibinfo {note} {arXiv:2208.05461}\BibitemShut {NoStop}%
\bibitem [{\citenamefont {Kang}\ \emph {et~al.}(2023)\citenamefont {Kang}, \citenamefont {Campbell},\ and\ \citenamefont {Brown}}]{kang_quantum_2023}%
  \BibitemOpen
  \bibfield  {author} {\bibinfo {author} {\bibfnamefont {M.}~\bibnamefont {Kang}}, \bibinfo {author} {\bibfnamefont {W.~C.}\ \bibnamefont {Campbell}}, \ and\ \bibinfo {author} {\bibfnamefont {K.~R.}\ \bibnamefont {Brown}},\ }\href {\doibase 10.1103/PRXQuantum.4.020358} {\bibfield  {journal} {\bibinfo  {journal} {PRX Quantum}\ }\textbf {\bibinfo {volume} {4}},\ \bibinfo {pages} {020358} (\bibinfo {year} {2023})}\BibitemShut {NoStop}%
\bibitem [{\citenamefont {Varbanov}\ \emph {et~al.}(2020)\citenamefont {Varbanov}, \citenamefont {Battistel}, \citenamefont {Tarasinski}, \citenamefont {Ostroukh}, \citenamefont {O’Brien}, \citenamefont {DiCarlo},\ and\ \citenamefont {Terhal}}]{varbanov_leakage_2020}%
  \BibitemOpen
  \bibfield  {author} {\bibinfo {author} {\bibfnamefont {B.~M.}\ \bibnamefont {Varbanov}}, \bibinfo {author} {\bibfnamefont {F.}~\bibnamefont {Battistel}}, \bibinfo {author} {\bibfnamefont {B.~M.}\ \bibnamefont {Tarasinski}}, \bibinfo {author} {\bibfnamefont {V.~P.}\ \bibnamefont {Ostroukh}}, \bibinfo {author} {\bibfnamefont {T.~E.}\ \bibnamefont {O’Brien}}, \bibinfo {author} {\bibfnamefont {L.}~\bibnamefont {DiCarlo}}, \ and\ \bibinfo {author} {\bibfnamefont {B.~M.}\ \bibnamefont {Terhal}},\ }\href {\doibase 10.1038/s41534-020-00330-w} {\bibfield  {journal} {\bibinfo  {journal} {npj Quantum Information}\ }\textbf {\bibinfo {volume} {6}},\ \bibinfo {pages} {1} (\bibinfo {year} {2020})}\BibitemShut {NoStop}%
\bibitem [{\citenamefont {Yang}\ \emph {et~al.}(2022)\citenamefont {Yang}, \citenamefont {Ma}, \citenamefont {Wu}, \citenamefont {Wang}, \citenamefont {Cao}, \citenamefont {Guo}, \citenamefont {Huang}, \citenamefont {Feng}, \citenamefont {Zhou},\ and\ \citenamefont {Duan}}]{yang_realizing_2022}%
  \BibitemOpen
  \bibfield  {author} {\bibinfo {author} {\bibfnamefont {H.-X.}\ \bibnamefont {Yang}}, \bibinfo {author} {\bibfnamefont {J.-Y.}\ \bibnamefont {Ma}}, \bibinfo {author} {\bibfnamefont {Y.-K.}\ \bibnamefont {Wu}}, \bibinfo {author} {\bibfnamefont {Y.}~\bibnamefont {Wang}}, \bibinfo {author} {\bibfnamefont {M.-M.}\ \bibnamefont {Cao}}, \bibinfo {author} {\bibfnamefont {W.-X.}\ \bibnamefont {Guo}}, \bibinfo {author} {\bibfnamefont {Y.-Y.}\ \bibnamefont {Huang}}, \bibinfo {author} {\bibfnamefont {L.}~\bibnamefont {Feng}}, \bibinfo {author} {\bibfnamefont {Z.-C.}\ \bibnamefont {Zhou}}, \ and\ \bibinfo {author} {\bibfnamefont {L.-M.}\ \bibnamefont {Duan}},\ }\href {\doibase 10.1038/s41567-022-01661-5} {\bibfield  {journal} {\bibinfo  {journal} {Nature Physics}\ }\textbf {\bibinfo {volume} {18}},\ \bibinfo {pages} {1058} (\bibinfo {year} {2022})}\BibitemShut {NoStop}%
\bibitem [{\citenamefont {Schmidt-Kaler}\ \emph {et~al.}(2003{\natexlab{a}})\citenamefont {Schmidt-Kaler}, \citenamefont {Häffner}, \citenamefont {Gulde}, \citenamefont {Riebe}, \citenamefont {Lancaster}, \citenamefont {Deuschle}, \citenamefont {Becher}, \citenamefont {Hänsel}, \citenamefont {Eschner}, \citenamefont {Roos},\ and\ \citenamefont {Blatt}}]{schmidt-kaler_how_2003}%
  \BibitemOpen
  \bibfield  {author} {\bibinfo {author} {\bibfnamefont {F.}~\bibnamefont {Schmidt-Kaler}}, \bibinfo {author} {\bibfnamefont {H.}~\bibnamefont {Häffner}}, \bibinfo {author} {\bibfnamefont {S.}~\bibnamefont {Gulde}}, \bibinfo {author} {\bibfnamefont {M.}~\bibnamefont {Riebe}}, \bibinfo {author} {\bibfnamefont {G.}~\bibnamefont {Lancaster}}, \bibinfo {author} {\bibfnamefont {T.}~\bibnamefont {Deuschle}}, \bibinfo {author} {\bibfnamefont {C.}~\bibnamefont {Becher}}, \bibinfo {author} {\bibfnamefont {W.}~\bibnamefont {Hänsel}}, \bibinfo {author} {\bibfnamefont {J.}~\bibnamefont {Eschner}}, \bibinfo {author} {\bibfnamefont {C.}~\bibnamefont {Roos}}, \ and\ \bibinfo {author} {\bibfnamefont {R.}~\bibnamefont {Blatt}},\ }\href {\doibase 10.1007/s00340-003-1346-9} {\bibfield  {journal} {\bibinfo  {journal} {Applied Physics B}\ }\textbf {\bibinfo {volume} {77}},\ \bibinfo {pages} {789} (\bibinfo {year} {2003}{\natexlab{a}})}\BibitemShut {NoStop}%
\bibitem [{\citenamefont {Schmidt-Kaler}\ \emph {et~al.}(2003{\natexlab{b}})\citenamefont {Schmidt-Kaler}, \citenamefont {Gulde}, \citenamefont {Riebe}, \citenamefont {Deuschle}, \citenamefont {Kreuter}, \citenamefont {Lancaster}, \citenamefont {Becher}, \citenamefont {Eschner}, \citenamefont {Häffner},\ and\ \citenamefont {Blatt}}]{schmidt-kaler_coherence_2003}%
  \BibitemOpen
  \bibfield  {author} {\bibinfo {author} {\bibfnamefont {F.}~\bibnamefont {Schmidt-Kaler}}, \bibinfo {author} {\bibfnamefont {S.}~\bibnamefont {Gulde}}, \bibinfo {author} {\bibfnamefont {M.}~\bibnamefont {Riebe}}, \bibinfo {author} {\bibfnamefont {T.}~\bibnamefont {Deuschle}}, \bibinfo {author} {\bibfnamefont {A.}~\bibnamefont {Kreuter}}, \bibinfo {author} {\bibfnamefont {G.}~\bibnamefont {Lancaster}}, \bibinfo {author} {\bibfnamefont {C.}~\bibnamefont {Becher}}, \bibinfo {author} {\bibfnamefont {J.}~\bibnamefont {Eschner}}, \bibinfo {author} {\bibfnamefont {H.}~\bibnamefont {Häffner}}, \ and\ \bibinfo {author} {\bibfnamefont {R.}~\bibnamefont {Blatt}},\ }\href {\doibase 10.1088/0953-4075/36/3/319} {\bibfield  {journal} {\bibinfo  {journal} {Journal of Physics B: Atomic, Molecular and Optical Physics}\ }\textbf {\bibinfo {volume} {36}},\ \bibinfo {pages} {623} (\bibinfo {year} {2003}{\natexlab{b}})}\BibitemShut {NoStop}%
\bibitem [{\citenamefont {Burrell}\ \emph {et~al.}(2010)\citenamefont {Burrell}, \citenamefont {Szwer}, \citenamefont {Webster},\ and\ \citenamefont {Lucas}}]{burrell_scalable_2010}%
  \BibitemOpen
  \bibfield  {author} {\bibinfo {author} {\bibfnamefont {A.~H.}\ \bibnamefont {Burrell}}, \bibinfo {author} {\bibfnamefont {D.~J.}\ \bibnamefont {Szwer}}, \bibinfo {author} {\bibfnamefont {S.~C.}\ \bibnamefont {Webster}}, \ and\ \bibinfo {author} {\bibfnamefont {D.~M.}\ \bibnamefont {Lucas}},\ }\href {\doibase 10.1103/PhysRevA.81.040302} {\bibfield  {journal} {\bibinfo  {journal} {Physical Review A}\ }\textbf {\bibinfo {volume} {81}},\ \bibinfo {pages} {040302} (\bibinfo {year} {2010})}\BibitemShut {NoStop}%
\bibitem [{\citenamefont {Gottesman}(1997)}]{gottesman_stabilizer_1997}%
  \BibitemOpen
  \bibfield  {author} {\bibinfo {author} {\bibfnamefont {D.}~\bibnamefont {Gottesman}},\ }\href {http://arxiv.org/abs/quant-ph/9705052} {\enquote {\bibinfo {title} {Stabilizer {Codes} and {Quantum} {Error} {Correction}},}\ } (\bibinfo {year} {1997}),\ \bibinfo {note} {arXiv:quant-ph/9705052}\BibitemShut {NoStop}%
\bibitem [{\citenamefont {Grassl}\ \emph {et~al.}(1997)\citenamefont {Grassl}, \citenamefont {Beth},\ and\ \citenamefont {Pellizzari}}]{grassl_codes_1997}%
  \BibitemOpen
  \bibfield  {author} {\bibinfo {author} {\bibfnamefont {M.}~\bibnamefont {Grassl}}, \bibinfo {author} {\bibfnamefont {T.}~\bibnamefont {Beth}}, \ and\ \bibinfo {author} {\bibfnamefont {T.}~\bibnamefont {Pellizzari}},\ }\href {\doibase 10.1103/PhysRevA.56.33} {\bibfield  {journal} {\bibinfo  {journal} {Physical Review A}\ }\textbf {\bibinfo {volume} {56}},\ \bibinfo {pages} {33} (\bibinfo {year} {1997})}\BibitemShut {NoStop}%
\bibitem [{\citenamefont {Bennett}\ \emph {et~al.}(1997)\citenamefont {Bennett}, \citenamefont {DiVincenzo},\ and\ \citenamefont {Smolin}}]{bennett_capacities_1997}%
  \BibitemOpen
  \bibfield  {author} {\bibinfo {author} {\bibfnamefont {C.~H.}\ \bibnamefont {Bennett}}, \bibinfo {author} {\bibfnamefont {D.~P.}\ \bibnamefont {DiVincenzo}}, \ and\ \bibinfo {author} {\bibfnamefont {J.~A.}\ \bibnamefont {Smolin}},\ }\href {\doibase 10.1103/PhysRevLett.78.3217} {\bibfield  {journal} {\bibinfo  {journal} {Physical Review Letters}\ }\textbf {\bibinfo {volume} {78}},\ \bibinfo {pages} {3217} (\bibinfo {year} {1997})}\BibitemShut {NoStop}%
\bibitem [{\citenamefont {Choonee}\ \emph {et~al.}(2017)\citenamefont {Choonee}, \citenamefont {Wilpers},\ and\ \citenamefont {Sinclair}}]{choonee_silicon_2017}%
  \BibitemOpen
  \bibfield  {author} {\bibinfo {author} {\bibfnamefont {K.}~\bibnamefont {Choonee}}, \bibinfo {author} {\bibfnamefont {G.}~\bibnamefont {Wilpers}}, \ and\ \bibinfo {author} {\bibfnamefont {A.~G.}\ \bibnamefont {Sinclair}},\ }in\ \href {\doibase 10.1109/TRANSDUCERS.2017.7994124} {\emph {\bibinfo {booktitle} {2017 19th {International} {Conference} on {Solid}-{State} {Sensors}, {Actuators} and {Microsystems} ({TRANSDUCERS})}}}\ (\bibinfo {year} {2017})\ pp.\ \bibinfo {pages} {615--618}\BibitemShut {NoStop}%
\bibitem [{\citenamefont {Wilpers}\ \emph {et~al.}(2013)\citenamefont {Wilpers}, \citenamefont {See}, \citenamefont {Gill},\ and\ \citenamefont {Sinclair}}]{wilpers_compact_2013}%
  \BibitemOpen
  \bibfield  {author} {\bibinfo {author} {\bibfnamefont {G.}~\bibnamefont {Wilpers}}, \bibinfo {author} {\bibfnamefont {P.}~\bibnamefont {See}}, \bibinfo {author} {\bibfnamefont {P.}~\bibnamefont {Gill}}, \ and\ \bibinfo {author} {\bibfnamefont {A.~G.}\ \bibnamefont {Sinclair}},\ }\href {\doibase 10.1007/s00340-012-5302-4} {\bibfield  {journal} {\bibinfo  {journal} {Applied Physics B}\ }\textbf {\bibinfo {volume} {111}},\ \bibinfo {pages} {21} (\bibinfo {year} {2013})}\BibitemShut {NoStop}%
\bibitem [{\citenamefont {Sotirova}\ \emph {et~al.}(2024)\citenamefont {Sotirova}, \citenamefont {Sun}, \citenamefont {Leppard}, \citenamefont {Wang}, \citenamefont {Wang}, \citenamefont {Vazquez-Brennan}, \citenamefont {Nadlinger}, \citenamefont {Moser}, \citenamefont {Jesacher}, \citenamefont {He}, \citenamefont {Pokorny}, \citenamefont {Booth},\ and\ \citenamefont {Ballance}}]{sotirova_low_2024}%
  \BibitemOpen
  \bibfield  {author} {\bibinfo {author} {\bibfnamefont {A.~S.}\ \bibnamefont {Sotirova}}, \bibinfo {author} {\bibfnamefont {B.}~\bibnamefont {Sun}}, \bibinfo {author} {\bibfnamefont {J.~D.}\ \bibnamefont {Leppard}}, \bibinfo {author} {\bibfnamefont {A.}~\bibnamefont {Wang}}, \bibinfo {author} {\bibfnamefont {M.}~\bibnamefont {Wang}}, \bibinfo {author} {\bibfnamefont {A.}~\bibnamefont {Vazquez-Brennan}}, \bibinfo {author} {\bibfnamefont {D.~P.}\ \bibnamefont {Nadlinger}}, \bibinfo {author} {\bibfnamefont {S.}~\bibnamefont {Moser}}, \bibinfo {author} {\bibfnamefont {A.}~\bibnamefont {Jesacher}}, \bibinfo {author} {\bibfnamefont {C.}~\bibnamefont {He}}, \bibinfo {author} {\bibfnamefont {F.}~\bibnamefont {Pokorny}}, \bibinfo {author} {\bibfnamefont {M.~J.}\ \bibnamefont {Booth}}, \ and\ \bibinfo {author} {\bibfnamefont {C.~J.}\ \bibnamefont {Ballance}},\ }\href {\doibase 10.1038/s41377-024-01542-x} {\bibfield  {journal} {\bibinfo  {journal} {Light: Science \& Applications}\ }\textbf {\bibinfo {volume} {13}},\
  \bibinfo {pages} {199} (\bibinfo {year} {2024})}\BibitemShut {NoStop}%
\bibitem [{\citenamefont {Sotirova}(2024)}]{sotirova_thesis_2024}%
  \BibitemOpen
  \bibfield  {author} {\bibinfo {author} {\bibfnamefont {A.~S.}\ \bibnamefont {Sotirova}},\ }\emph {\bibinfo {title} {Trapped Ion Quantum Information Processing Using Multiple Qubit Encodings}},\ \href@noop {} {Ph.D. thesis},\ \bibinfo  {school} {University of Oxford, UK} (\bibinfo {year} {2024})\BibitemShut {NoStop}%
\bibitem [{\citenamefont {Leupold}\ \emph {et~al.}(2018)\citenamefont {Leupold}, \citenamefont {Malinowski}, \citenamefont {Zhang}, \citenamefont {Negnevitsky}, \citenamefont {Cabello}, \citenamefont {Alonso},\ and\ \citenamefont {Home}}]{leupold_sustained_2018}%
  \BibitemOpen
  \bibfield  {author} {\bibinfo {author} {\bibfnamefont {F.}~\bibnamefont {Leupold}}, \bibinfo {author} {\bibfnamefont {M.}~\bibnamefont {Malinowski}}, \bibinfo {author} {\bibfnamefont {C.}~\bibnamefont {Zhang}}, \bibinfo {author} {\bibfnamefont {V.}~\bibnamefont {Negnevitsky}}, \bibinfo {author} {\bibfnamefont {A.}~\bibnamefont {Cabello}}, \bibinfo {author} {\bibfnamefont {J.}~\bibnamefont {Alonso}}, \ and\ \bibinfo {author} {\bibfnamefont {J.}~\bibnamefont {Home}},\ }\href {\doibase 10.1103/PhysRevLett.120.180401} {\bibfield  {journal} {\bibinfo  {journal} {Physical Review Letters}\ }\textbf {\bibinfo {volume} {120}},\ \bibinfo {pages} {180401} (\bibinfo {year} {2018})}\BibitemShut {NoStop}%
\bibitem [{\citenamefont {Szwer}(2009)}]{szwer_high_2009}%
  \BibitemOpen
  \bibfield  {author} {\bibinfo {author} {\bibfnamefont {D.}~\bibnamefont {Szwer}},\ }\emph {\bibinfo {title} {High fidelity readout and protection of a $^{43}Ca^+$ trapped ion qubit}},\ \href@noop {} {Ph.D. thesis},\ \bibinfo  {school} {University of Oxford, UK} (\bibinfo {year} {2009})\BibitemShut {NoStop}%
\bibitem [{\citenamefont {Neuhauser}\ \emph {et~al.}(1978)\citenamefont {Neuhauser}, \citenamefont {Hohenstatt}, \citenamefont {Toschek},\ and\ \citenamefont {Dehmelt}}]{neuhauser_optical-sideband_1978}%
  \BibitemOpen
  \bibfield  {author} {\bibinfo {author} {\bibfnamefont {W.}~\bibnamefont {Neuhauser}}, \bibinfo {author} {\bibfnamefont {M.}~\bibnamefont {Hohenstatt}}, \bibinfo {author} {\bibfnamefont {P.}~\bibnamefont {Toschek}}, \ and\ \bibinfo {author} {\bibfnamefont {H.}~\bibnamefont {Dehmelt}},\ }\href {\doibase 10.1103/PhysRevLett.41.233} {\bibfield  {journal} {\bibinfo  {journal} {Physical Review Letters}\ }\textbf {\bibinfo {volume} {41}},\ \bibinfo {pages} {233} (\bibinfo {year} {1978})}\BibitemShut {NoStop}%
\bibitem [{\citenamefont {Wineland}\ \emph {et~al.}(1978)\citenamefont {Wineland}, \citenamefont {Drullinger},\ and\ \citenamefont {Walls}}]{wineland_radiation-pressure_1978}%
  \BibitemOpen
  \bibfield  {author} {\bibinfo {author} {\bibfnamefont {D.~J.}\ \bibnamefont {Wineland}}, \bibinfo {author} {\bibfnamefont {R.~E.}\ \bibnamefont {Drullinger}}, \ and\ \bibinfo {author} {\bibfnamefont {F.~L.}\ \bibnamefont {Walls}},\ }\href {\doibase 10.1103/PhysRevLett.40.1639} {\bibfield  {journal} {\bibinfo  {journal} {Physical Review Letters}\ }\textbf {\bibinfo {volume} {40}},\ \bibinfo {pages} {1639} (\bibinfo {year} {1978})}\BibitemShut {NoStop}%
\bibitem [{\citenamefont {Wineland}\ and\ \citenamefont {Itano}(1979)}]{wineland_laser_1979}%
  \BibitemOpen
  \bibfield  {author} {\bibinfo {author} {\bibfnamefont {D.~J.}\ \bibnamefont {Wineland}}\ and\ \bibinfo {author} {\bibfnamefont {W.~M.}\ \bibnamefont {Itano}},\ }\href {\doibase 10.1103/PhysRevA.20.1521} {\bibfield  {journal} {\bibinfo  {journal} {Physical Review A}\ }\textbf {\bibinfo {volume} {20}},\ \bibinfo {pages} {1521} (\bibinfo {year} {1979})}\BibitemShut {NoStop}%
\bibitem [{\citenamefont {Burrell}(2010)}]{burrell_high_2010}%
  \BibitemOpen
  \bibfield  {author} {\bibinfo {author} {\bibfnamefont {A.}~\bibnamefont {Burrell}},\ }\emph {\bibinfo {title} {High fidelity readout of trapped ion qubits}},\ \href@noop {} {Ph.D. thesis},\ \bibinfo  {school} {University of Oxford, UK} (\bibinfo {year} {2010})\BibitemShut {NoStop}%
\bibitem [{\citenamefont {Leung}\ \emph {et~al.}(1997)\citenamefont {Leung}, \citenamefont {Nielsen}, \citenamefont {Chuang},\ and\ \citenamefont {Yamamoto}}]{leung_approximate_1997}%
  \BibitemOpen
  \bibfield  {author} {\bibinfo {author} {\bibfnamefont {D.~W.}\ \bibnamefont {Leung}}, \bibinfo {author} {\bibfnamefont {M.~A.}\ \bibnamefont {Nielsen}}, \bibinfo {author} {\bibfnamefont {I.~L.}\ \bibnamefont {Chuang}}, \ and\ \bibinfo {author} {\bibfnamefont {Y.}~\bibnamefont {Yamamoto}},\ }\href {\doibase 10.1103/PhysRevA.56.2567} {\bibfield  {journal} {\bibinfo  {journal} {Physical Review A}\ }\textbf {\bibinfo {volume} {56}},\ \bibinfo {pages} {2567} (\bibinfo {year} {1997})}\BibitemShut {NoStop}%
\bibitem [{\citenamefont {Itano}\ \emph {et~al.}(1993)\citenamefont {Itano}, \citenamefont {Bergquist}, \citenamefont {Bollinger}, \citenamefont {Gilligan}, \citenamefont {Heinzen}, \citenamefont {Moore}, \citenamefont {Raizen},\ and\ \citenamefont {Wineland}}]{itano_quantum_1993}%
  \BibitemOpen
  \bibfield  {author} {\bibinfo {author} {\bibfnamefont {W.~M.}\ \bibnamefont {Itano}}, \bibinfo {author} {\bibfnamefont {J.~C.}\ \bibnamefont {Bergquist}}, \bibinfo {author} {\bibfnamefont {J.~J.}\ \bibnamefont {Bollinger}}, \bibinfo {author} {\bibfnamefont {J.~M.}\ \bibnamefont {Gilligan}}, \bibinfo {author} {\bibfnamefont {D.~J.}\ \bibnamefont {Heinzen}}, \bibinfo {author} {\bibfnamefont {F.~L.}\ \bibnamefont {Moore}}, \bibinfo {author} {\bibfnamefont {M.~G.}\ \bibnamefont {Raizen}}, \ and\ \bibinfo {author} {\bibfnamefont {D.~J.}\ \bibnamefont {Wineland}},\ }\href {\doibase 10.1103/PhysRevA.47.3554} {\bibfield  {journal} {\bibinfo  {journal} {Physical Review A}\ }\textbf {\bibinfo {volume} {47}},\ \bibinfo {pages} {3554} (\bibinfo {year} {1993})}\BibitemShut {NoStop}%
\bibitem [{\citenamefont {Cao}\ \emph {et~al.}(2018)\citenamefont {Cao}, \citenamefont {Romero},\ and\ \citenamefont {Aspuru-Guzik}}]{cao_potential_2018}%
  \BibitemOpen
  \bibfield  {author} {\bibinfo {author} {\bibfnamefont {Y.}~\bibnamefont {Cao}}, \bibinfo {author} {\bibfnamefont {J.}~\bibnamefont {Romero}}, \ and\ \bibinfo {author} {\bibfnamefont {A.}~\bibnamefont {Aspuru-Guzik}},\ }\href {\doibase 10.1147/JRD.2018.2888987} {\bibfield  {journal} {\bibinfo  {journal} {IBM Journal of Research and Development}\ }\textbf {\bibinfo {volume} {62}},\ \bibinfo {pages} {6:1} (\bibinfo {year} {2018})}\BibitemShut {NoStop}%
\bibitem [{\citenamefont {Cerezo}\ \emph {et~al.}(2021)\citenamefont {Cerezo}, \citenamefont {Arrasmith}, \citenamefont {Babbush}, \citenamefont {Benjamin}, \citenamefont {Endo}, \citenamefont {Fujii}, \citenamefont {McClean}, \citenamefont {Mitarai}, \citenamefont {Yuan}, \citenamefont {Cincio},\ and\ \citenamefont {Coles}}]{cerezo_variational_2021}%
  \BibitemOpen
  \bibfield  {author} {\bibinfo {author} {\bibfnamefont {M.}~\bibnamefont {Cerezo}}, \bibinfo {author} {\bibfnamefont {A.}~\bibnamefont {Arrasmith}}, \bibinfo {author} {\bibfnamefont {R.}~\bibnamefont {Babbush}}, \bibinfo {author} {\bibfnamefont {S.~C.}\ \bibnamefont {Benjamin}}, \bibinfo {author} {\bibfnamefont {S.}~\bibnamefont {Endo}}, \bibinfo {author} {\bibfnamefont {K.}~\bibnamefont {Fujii}}, \bibinfo {author} {\bibfnamefont {J.~R.}\ \bibnamefont {McClean}}, \bibinfo {author} {\bibfnamefont {K.}~\bibnamefont {Mitarai}}, \bibinfo {author} {\bibfnamefont {X.}~\bibnamefont {Yuan}}, \bibinfo {author} {\bibfnamefont {L.}~\bibnamefont {Cincio}}, \ and\ \bibinfo {author} {\bibfnamefont {P.~J.}\ \bibnamefont {Coles}},\ }\href {\doibase 10.1038/s42254-021-00348-9} {\bibfield  {journal} {\bibinfo  {journal} {Nature Reviews Physics}\ }\textbf {\bibinfo {volume} {3}},\ \bibinfo {pages} {625} (\bibinfo {year} {2021})}\BibitemShut {NoStop}%
\bibitem [{\citenamefont {Ferguson}\ \emph {et~al.}(2021)\citenamefont {Ferguson}, \citenamefont {Dellantonio}, \citenamefont {Balushi}, \citenamefont {Jansen}, \citenamefont {Dür},\ and\ \citenamefont {Muschik}}]{ferguson_measurement-based_2021}%
  \BibitemOpen
  \bibfield  {author} {\bibinfo {author} {\bibfnamefont {R.}~\bibnamefont {Ferguson}}, \bibinfo {author} {\bibfnamefont {L.}~\bibnamefont {Dellantonio}}, \bibinfo {author} {\bibfnamefont {A.~A.}\ \bibnamefont {Balushi}}, \bibinfo {author} {\bibfnamefont {K.}~\bibnamefont {Jansen}}, \bibinfo {author} {\bibfnamefont {W.}~\bibnamefont {Dür}}, \ and\ \bibinfo {author} {\bibfnamefont {C.}~\bibnamefont {Muschik}},\ }\href {\doibase 10.1103/PhysRevLett.126.220501} {\bibfield  {journal} {\bibinfo  {journal} {Physical Review Letters}\ }\textbf {\bibinfo {volume} {126}},\ \bibinfo {pages} {220501} (\bibinfo {year} {2021})}\BibitemShut {NoStop}%
\bibitem [{\citenamefont {Astafiev}\ \emph {et~al.}(2010)\citenamefont {Astafiev}, \citenamefont {Zagoskin}, \citenamefont {Abdumalikov}, \citenamefont {Pashkin}, \citenamefont {Yamamoto}, \citenamefont {Inomata}, \citenamefont {Nakamura},\ and\ \citenamefont {Tsai}}]{astafiev_resonance_2010}%
  \BibitemOpen
  \bibfield  {author} {\bibinfo {author} {\bibfnamefont {O.}~\bibnamefont {Astafiev}}, \bibinfo {author} {\bibfnamefont {A.~M.}\ \bibnamefont {Zagoskin}}, \bibinfo {author} {\bibfnamefont {A.~A.}\ \bibnamefont {Abdumalikov}}, \bibinfo {author} {\bibfnamefont {Y.~A.}\ \bibnamefont {Pashkin}}, \bibinfo {author} {\bibfnamefont {T.}~\bibnamefont {Yamamoto}}, \bibinfo {author} {\bibfnamefont {K.}~\bibnamefont {Inomata}}, \bibinfo {author} {\bibfnamefont {Y.}~\bibnamefont {Nakamura}}, \ and\ \bibinfo {author} {\bibfnamefont {J.~S.}\ \bibnamefont {Tsai}},\ }\href {\doibase 10.1126/science.1181918} {\bibfield  {journal} {\bibinfo  {journal} {Science}\ }\textbf {\bibinfo {volume} {327}},\ \bibinfo {pages} {840} (\bibinfo {year} {2010})}\BibitemShut {NoStop}%
\bibitem [{\citenamefont {Cottet}\ \emph {et~al.}(2021)\citenamefont {Cottet}, \citenamefont {Xiong}, \citenamefont {Nguyen}, \citenamefont {Lin},\ and\ \citenamefont {Manucharyan}}]{cottet_electron_2021}%
  \BibitemOpen
  \bibfield  {author} {\bibinfo {author} {\bibfnamefont {N.}~\bibnamefont {Cottet}}, \bibinfo {author} {\bibfnamefont {H.}~\bibnamefont {Xiong}}, \bibinfo {author} {\bibfnamefont {L.~B.}\ \bibnamefont {Nguyen}}, \bibinfo {author} {\bibfnamefont {Y.-H.}\ \bibnamefont {Lin}}, \ and\ \bibinfo {author} {\bibfnamefont {V.~E.}\ \bibnamefont {Manucharyan}},\ }\href {\doibase 10.1038/s41467-021-26686-x} {\bibfield  {journal} {\bibinfo  {journal} {Nature Communications}\ }\textbf {\bibinfo {volume} {12}},\ \bibinfo {pages} {6383} (\bibinfo {year} {2021})}\BibitemShut {NoStop}%
\bibitem [{\citenamefont {Low}\ \emph {et~al.}(2020)\citenamefont {Low}, \citenamefont {White}, \citenamefont {Cox}, \citenamefont {Day},\ and\ \citenamefont {Senko}}]{low_practical_2020}%
  \BibitemOpen
  \bibfield  {author} {\bibinfo {author} {\bibfnamefont {P.~J.}\ \bibnamefont {Low}}, \bibinfo {author} {\bibfnamefont {B.~M.}\ \bibnamefont {White}}, \bibinfo {author} {\bibfnamefont {A.~A.}\ \bibnamefont {Cox}}, \bibinfo {author} {\bibfnamefont {M.~L.}\ \bibnamefont {Day}}, \ and\ \bibinfo {author} {\bibfnamefont {C.}~\bibnamefont {Senko}},\ }\href {\doibase 10.1103/PhysRevResearch.2.033128} {\bibfield  {journal} {\bibinfo  {journal} {Physical Review Research}\ }\textbf {\bibinfo {volume} {2}},\ \bibinfo {pages} {033128} (\bibinfo {year} {2020})}\BibitemShut {NoStop}%
\bibitem [{\citenamefont {Ringbauer}\ \emph {et~al.}(2022)\citenamefont {Ringbauer}, \citenamefont {Meth}, \citenamefont {Postler}, \citenamefont {Stricker}, \citenamefont {Blatt}, \citenamefont {Schindler},\ and\ \citenamefont {Monz}}]{ringbauer_universal_2022}%
  \BibitemOpen
  \bibfield  {author} {\bibinfo {author} {\bibfnamefont {M.}~\bibnamefont {Ringbauer}}, \bibinfo {author} {\bibfnamefont {M.}~\bibnamefont {Meth}}, \bibinfo {author} {\bibfnamefont {L.}~\bibnamefont {Postler}}, \bibinfo {author} {\bibfnamefont {R.}~\bibnamefont {Stricker}}, \bibinfo {author} {\bibfnamefont {R.}~\bibnamefont {Blatt}}, \bibinfo {author} {\bibfnamefont {P.}~\bibnamefont {Schindler}}, \ and\ \bibinfo {author} {\bibfnamefont {T.}~\bibnamefont {Monz}},\ }\href {\doibase 10.1038/s41567-022-01658-0} {\bibfield  {journal} {\bibinfo  {journal} {Nature Physics}\ }\textbf {\bibinfo {volume} {18}},\ \bibinfo {pages} {1053} (\bibinfo {year} {2022})}\BibitemShut {NoStop}%
\bibitem [{\citenamefont {Low}\ \emph {et~al.}(2023)\citenamefont {Low}, \citenamefont {White},\ and\ \citenamefont {Senko}}]{low_control_2023}%
  \BibitemOpen
  \bibfield  {author} {\bibinfo {author} {\bibfnamefont {P.~J.}\ \bibnamefont {Low}}, \bibinfo {author} {\bibfnamefont {B.}~\bibnamefont {White}}, \ and\ \bibinfo {author} {\bibfnamefont {C.}~\bibnamefont {Senko}},\ }\href {\doibase 10.48550/arXiv.2306.03340} {\enquote {\bibinfo {title} {Control and {Readout} of a 13-level {Trapped} {Ion} {Qudit}},}\ } (\bibinfo {year} {2023}),\ \bibinfo {note} {arXiv:2306.03340}\BibitemShut {NoStop}%
\bibitem [{\citenamefont {Hrmo}\ \emph {et~al.}(2023)\citenamefont {Hrmo}, \citenamefont {Wilhelm}, \citenamefont {Gerster}, \citenamefont {van Mourik}, \citenamefont {Huber}, \citenamefont {Blatt}, \citenamefont {Schindler}, \citenamefont {Monz},\ and\ \citenamefont {Ringbauer}}]{hrmo_native_2023}%
  \BibitemOpen
  \bibfield  {author} {\bibinfo {author} {\bibfnamefont {P.}~\bibnamefont {Hrmo}}, \bibinfo {author} {\bibfnamefont {B.}~\bibnamefont {Wilhelm}}, \bibinfo {author} {\bibfnamefont {L.}~\bibnamefont {Gerster}}, \bibinfo {author} {\bibfnamefont {M.~W.}\ \bibnamefont {van Mourik}}, \bibinfo {author} {\bibfnamefont {M.}~\bibnamefont {Huber}}, \bibinfo {author} {\bibfnamefont {R.}~\bibnamefont {Blatt}}, \bibinfo {author} {\bibfnamefont {P.}~\bibnamefont {Schindler}}, \bibinfo {author} {\bibfnamefont {T.}~\bibnamefont {Monz}}, \ and\ \bibinfo {author} {\bibfnamefont {M.}~\bibnamefont {Ringbauer}},\ }\href {\doibase 10.1038/s41467-023-37375-2} {\bibfield  {journal} {\bibinfo  {journal} {Nature Communications}\ }\textbf {\bibinfo {volume} {14}},\ \bibinfo {pages} {2242} (\bibinfo {year} {2023})}\BibitemShut {NoStop}%
\bibitem [{\citenamefont {Goss}\ \emph {et~al.}(2023)\citenamefont {Goss}, \citenamefont {Ferracin}, \citenamefont {Hashim}, \citenamefont {Carignan-Dugas}, \citenamefont {Kreikebaum}, \citenamefont {Naik}, \citenamefont {Santiago},\ and\ \citenamefont {Siddiqi}}]{goss_extending_2023}%
  \BibitemOpen
  \bibfield  {author} {\bibinfo {author} {\bibfnamefont {N.}~\bibnamefont {Goss}}, \bibinfo {author} {\bibfnamefont {S.}~\bibnamefont {Ferracin}}, \bibinfo {author} {\bibfnamefont {A.}~\bibnamefont {Hashim}}, \bibinfo {author} {\bibfnamefont {A.}~\bibnamefont {Carignan-Dugas}}, \bibinfo {author} {\bibfnamefont {J.~M.}\ \bibnamefont {Kreikebaum}}, \bibinfo {author} {\bibfnamefont {R.~K.}\ \bibnamefont {Naik}}, \bibinfo {author} {\bibfnamefont {D.~I.}\ \bibnamefont {Santiago}}, \ and\ \bibinfo {author} {\bibfnamefont {I.}~\bibnamefont {Siddiqi}},\ }\href {\doibase 10.48550/arXiv.2305.16507} {\enquote {\bibinfo {title} {Extending the {Computational} {Reach} of a {Superconducting} {Qutrit} {Processor}},}\ } (\bibinfo {year} {2023}),\ \bibinfo {note} {arXiv:2305.16507}\BibitemShut {NoStop}%
\bibitem [{\citenamefont {Leibfried}(2012)}]{leibfried_quantum_2012}%
  \BibitemOpen
  \bibfield  {author} {\bibinfo {author} {\bibfnamefont {D.}~\bibnamefont {Leibfried}},\ }\href {\doibase 10.1088/1367-2630/14/2/023029} {\bibfield  {journal} {\bibinfo  {journal} {New Journal of Physics}\ }\textbf {\bibinfo {volume} {14}},\ \bibinfo {pages} {023029} (\bibinfo {year} {2012})}\BibitemShut {NoStop}%
\bibitem [{\citenamefont {Wolf}\ \emph {et~al.}(2024)\citenamefont {Wolf}, \citenamefont {Heip}, \citenamefont {Zawierucha}, \citenamefont {Shi}, \citenamefont {Ospelkaus},\ and\ \citenamefont {Schmidt}}]{wolf_prospect_2024}%
  \BibitemOpen
  \bibfield  {author} {\bibinfo {author} {\bibfnamefont {F.}~\bibnamefont {Wolf}}, \bibinfo {author} {\bibfnamefont {J.~C.}\ \bibnamefont {Heip}}, \bibinfo {author} {\bibfnamefont {M.~J.}\ \bibnamefont {Zawierucha}}, \bibinfo {author} {\bibfnamefont {C.}~\bibnamefont {Shi}}, \bibinfo {author} {\bibfnamefont {S.}~\bibnamefont {Ospelkaus}}, \ and\ \bibinfo {author} {\bibfnamefont {P.~O.}\ \bibnamefont {Schmidt}},\ }\href {\doibase 10.1088/1367-2630/ad1ad3} {\bibfield  {journal} {\bibinfo  {journal} {New Journal of Physics}\ }\textbf {\bibinfo {volume} {26}},\ \bibinfo {pages} {013028} (\bibinfo {year} {2024})}\BibitemShut {NoStop}%
\bibitem [{\citenamefont {Zhang}\ \emph {et~al.}(2020)\citenamefont {Zhang}, \citenamefont {Arnold}, \citenamefont {Chanu}, \citenamefont {Kaewuam}, \citenamefont {Safronova},\ and\ \citenamefont {Barrett}}]{zhang_branching_2020}%
  \BibitemOpen
  \bibfield  {author} {\bibinfo {author} {\bibfnamefont {Z.}~\bibnamefont {Zhang}}, \bibinfo {author} {\bibfnamefont {K.~J.}\ \bibnamefont {Arnold}}, \bibinfo {author} {\bibfnamefont {S.~R.}\ \bibnamefont {Chanu}}, \bibinfo {author} {\bibfnamefont {R.}~\bibnamefont {Kaewuam}}, \bibinfo {author} {\bibfnamefont {M.~S.}\ \bibnamefont {Safronova}}, \ and\ \bibinfo {author} {\bibfnamefont {M.~D.}\ \bibnamefont {Barrett}},\ }\href {\doibase 10.1103/PhysRevA.101.062515} {\bibfield  {journal} {\bibinfo  {journal} {Physical Review A}\ }\textbf {\bibinfo {volume} {101}},\ \bibinfo {pages} {062515} (\bibinfo {year} {2020})}\BibitemShut {NoStop}%
\end{thebibliography}%

\pagebreak

\widetext
\clearpage
\begin{center}

\end{center}

\setcounter{secnumdepth}{4}
\setcounter{equation}{0}
\setcounter{figure}{0}
\setcounter{table}{0}
\setcounter{section}{0}
\setcounter{page}{1}
\makeatletter
\renewcommand{\theequation}{S\arabic{equation}}
\renewcommand{\thefigure}{S\arabic{figure}}
\renewcommand{\thesection}{S\arabic{section}}

\section{Barium level structure and basic laser operations}
\label{sec:si-ba-level-structure}
\begin{figure}[!ht]
    \centering
    \includegraphics{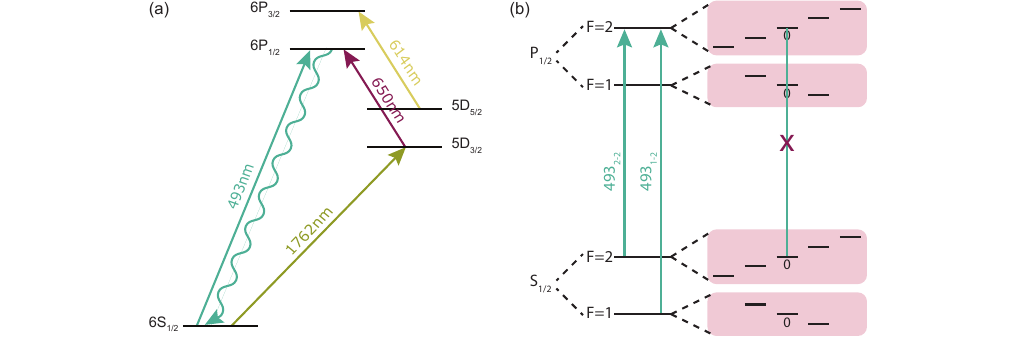}
    \caption{(a) An overview of level structure of \ba. The ions are cooled and detected by repeatedly exciting the transition between the \ground and the \pone levels and collecting the scattered $493\nm$ light. During that period, $650\nm$ laser is used to repump population from the \dthree level. The $1762\nm$ laser is used to transfer population to and from the metastable \shelf level. The $614\nm$ laser removes population from the \shelf level at the end of the experiment. (b) State initialization of \ba using optical pumping. The $493\nm$ laser is frequency-modulated by an EOM to only turn on the sidebands resonant with the $\ket{S_{1/2}, F=2}\leftrightarrow\ket{P_{1/2}, F=2}$ and $\ket{S_{1/2}, F=1}\leftrightarrow\ket{P_{1/2}, F=2}$ transitions. Since the $\ket{S_{1/2}, F=2, m_F=0}\leftrightarrow\ket{P_{1/2}, F=2, m_F=0}$ transition is forbidden, this process results in the majority of the population being pumped in the $\ket{S_{1/2}, F=2, m_F=0}$ state. Further details can be found in Ref.~\cite{sotirova_thesis_2024}.}
    \label{fig:ba-levels-state-prep}
\end{figure}

\begin{figure}[!ht]
    \centering
    \includegraphics{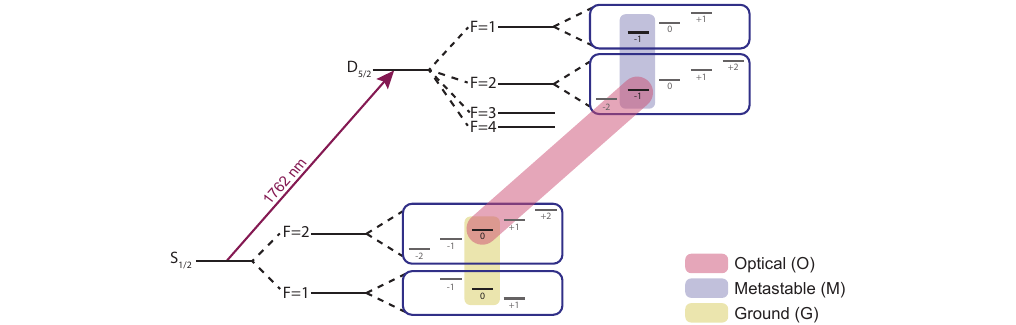}
    \caption{The pairs of states used as O, M, and G qubits in our experiment. The states in black are the qubit states. The states in grey are used as intermediate states for population transfer in the SPAM sequence.}
    \label{fig:qubit-states}
\end{figure}
\newpage
\section{Experiment setup}
\label{sec:si-exp-setup}
\begin{figure}[!ht]
    \centering
    \includegraphics{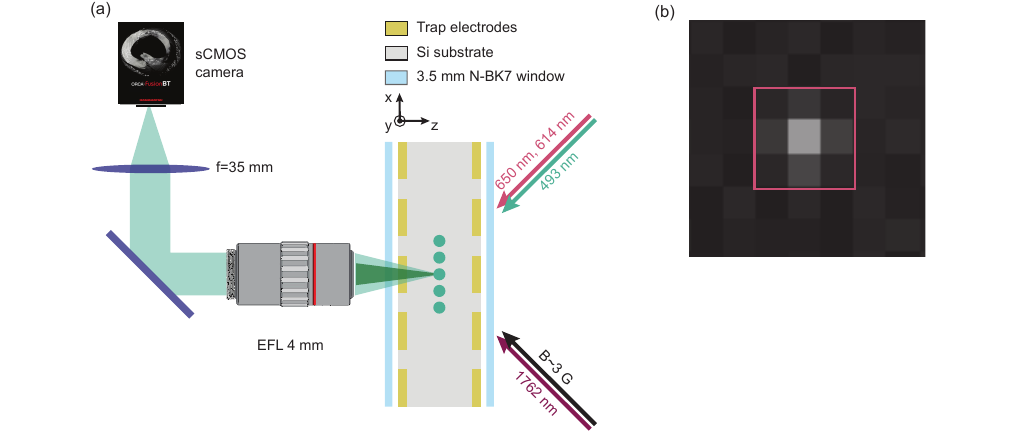}
    \caption{Experiment setup. (a) The \ba ion is confined in a segmented 3D monolithic trap \cite{choonee_silicon_2017, wilpers_compact_2013}. The $493\nm$ light scattered by the ion is collimated using a commercial $\mathrm{NA}0.5$ microscope objective (Mitotoyo G Plan Apo 50) with an effective focal length (EFL) of $4\mm$ and refocused on an sCMOS camera (Hamamatsu ORCA Fusion-BT) using an $f=35\mm$ lens for spatially resolved readout. The applied magnetic field of $\approx 3\,\mathrm{G}$ is used to lift the degeneracy of the Zeeman states. The laser beams incident from the right are used for ion cooling, state preparation, and measurement, as well as for population transfer to and from the metastable \shelf level [see fig.~\ref{fig:ba-levels-state-prep}(a)]. (b) An image of a single trapped \ba ion taken with the setup shown in (a). To determine the amount of fluorescence from the ion, we sum the counts from the pixels within a $3\times3$ square around the center position of the ion. Further details can be found in Ref.~\cite{sotirova_thesis_2024}.}
    \label{fig:exp-setup}
\end{figure}

\section{Detailed SPAM sequence}
\label{sec:si-spam-sequence}
In this section we provide the detailed SPAM steps and the SPAM measurement data for each qubit type in \ba.

\subsection{Optical qubit}
\label{sec:si-spam-steps-optical}
\begin{enumerate}
    \item Population detection step $R_0$.
    \item Optical pumping into $\ket{\groundm, F=2, m_F=0}$.
    \item Population transfer using a $1762\nm$ pulse from $\ket{\groundm, F=2, m_F=0}$ to either $\ket{0}=\ket{\shelfm, F=2, m_F=-1}$ or $\ket{\shelfm, F=1, m_F=-1}$.
    \item Population detection step $R_1$.
    \item If preparing $\ket{1}=\ket{\groundm, F=2, m_F=0}$, population transfer using a $1762\nm$ pulse from $\ket{\shelfm, F=1, m_F=-1}$ to $\ket{\groundm, F=2, m_F=0}$.
    \item If preparing $\ket{1}$, population transfer using a $1762\nm$ pulse from the state $\ket{\groundm, F=2, m_F=0}$ to $\ket{\shelfm, F=1, m_F=-1}$.
    \item Population detection step $R_2$.
    \item Population transfer using a $1762\nm$ pulse from $\ket{\shelfm, F=2, m_F=-1}$ to $\ket{\groundm, F=2, m_F=0}$.
    \item Population detection step $R_3$.
    \item Population transfer using a $1762\nm$ pulse from $\ket{\shelfm, F=1, m_F=-1}$ to $\ket{\groundm, F=2, m_F=0}$.
    \item Population detection step $R_4$.
    \item Population detection step $R_5$.
\end{enumerate}

\begin{figure}[!ht]
    \centering
    \includegraphics{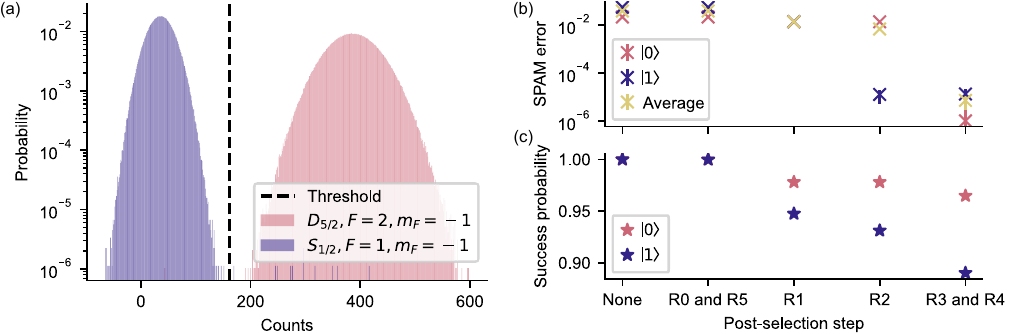}
    \caption{Optical qubit SPAM data. The SPAM error after the final post-selection step is $(7\pm4)\times10^{-6}$.}
    \label{fig:si-optical-spam}
\end{figure}

\subsection{Metastable level qubit}
\label{sec:si-spam-steps-metastable}
\begin{enumerate}
    \item Population detection step $R_0$.
    \item Optical pumping into $\ket{\groundm, F=2, m_F=0}$.
    \item Population transfer using a $1762\nm$ pulse from $\ket{\groundm, F=2, m_F=0}$ to either $\ket{0}=\ket{\shelfm, F=2, m_F=-1}$ or $\ket{1}=\ket{\shelfm, F=1, m_F=-1}$.
    \item Population detection step $R_1$.
    \item Population detection step $R_2$.
    \item Population transfer using a $1762\nm$ pulse from $\ket{\shelfm, F=2, m_F=-1}$ to $\ket{\groundm, F=2, m_F=0}$.
    \item Population detection step $R_3$.
    \item Population transfer using a $1762\nm$ pulse from $\ket{\shelfm, F=1, m_F=-1}$ to $\ket{\groundm, F=2, m_F=0}$.
    \item Population detection step $R_4$.
    \item Population detection step $R_5$.
\end{enumerate}

\begin{figure}[!ht]
    \centering
    \includegraphics{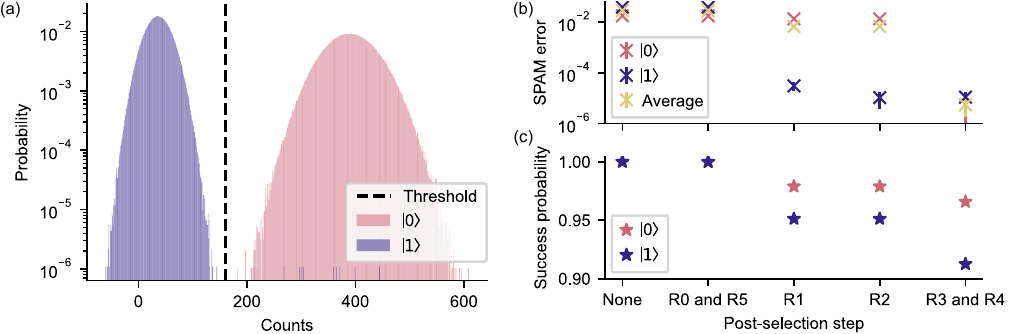}
    \caption{Metastable level qubit SPAM data (presented here in addition to fig.~\ref{fig:spam-data-metastable} in the main text for completeness). The SPAM error after the final post-selection step is $(5\pm4)\times10^{-6}$.}
    \label{fig:si-metastable-spam}
\end{figure}

\subsection{Ground level qubit}
\label{sec:si-spam-steps-ground}
\begin{enumerate}
    \item Population detection step $R_0$.
    \item Optical pumping into $\ket{\groundm, F=2, m_F=0}$.
    \item Population transfer using a $1762\nm$ pulse from $\ket{\groundm, F=2, m_F=0}$ to $\ket{\shelfm, F=2, m_F=-1}$.
    \item Population detection step $R_1$.
    \item Population transfer using a $1762\nm$ pulse from $\ket{\shelfm, F=2, m_F=-1}$ to either $\ket{0}=\ket{\groundm, F=2, m_F=0}$ or $\ket{1}=\ket{\groundm, F=1, m_F=0}$.
    \item Population transfer using a $1762\nm$ pulse from $\ket{\groundm, F=2, m_F=0}$ to $\ket{\shelfm, F=2, m_F=+1}$.
    \item Population transfer using a $1762\nm$ pulse from $\ket{\groundm, F=1, m_F=0}$ to $\ket{D_{5/2}, F=1, m_F=-1}$.
    \item Population detection step $R_2$.
    \item Population transfer using a $1762\nm$ pulse from $\ket{\shelfm, F=2, m_F=+1}$ to $\ket{\groundm, F=2, m_F=0}$.
    \item Population detection step $R_3$.
    \item Population transfer using a $1762\nm$ pulse from $\ket{\shelfm, F=1, m_F=-1}$ to $\ket{\groundm, F=2, m_F=0}$.
    \item Population detection step $R_4$.
    \item Population detection step $R_5$.
\end{enumerate}

\begin{figure}[!ht]
    \centering
    \includegraphics{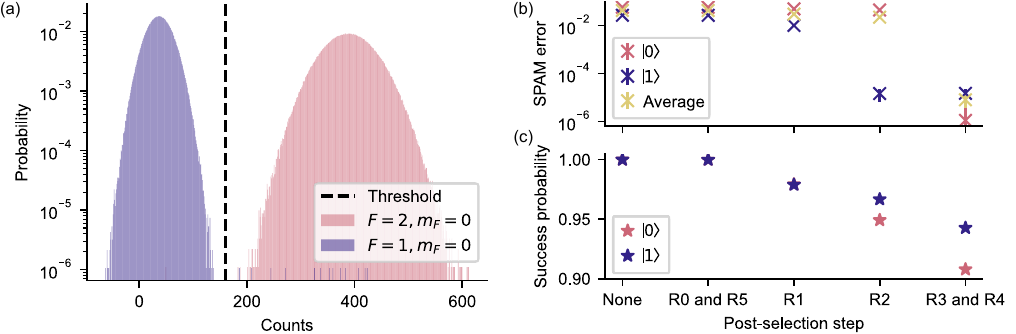}
    \caption{Ground level qubit SPAM data. The SPAM error after the final post-selection step is $(8\pm4)\times10^{-6}$.}
    \label{fig:si-ground-spam}
\end{figure}

\section{SPAM error budget}
\label{sec:si-error-budget}
In this section, we present the error budget for the SPAM protocol. It is structured as follows. First, we define the population detection error and measure its two components: the optical detection error and the metastable level decay probability. Then, we calculate the total population detection error and argue why the total error of the SPAM protocol is equal to the population detection error. Further details on the concepts and measurements presented in this section can be found in Ref.~\cite{sotirova_thesis_2024}.

In an ideal population detection, an ion that begins in \ground is recorded as bright, and an ion that begins in \shelf is recorded as dark. There are two physical sources from which a population detection error can arise. The first -- what we call an ``optical detection error'' -- arises due to the statistics of photon counts collected from ions in the \ground and \shelf levels. The second arises because the ion can spontaneously decay from \shelf to \ground, at which point it begins to fluoresce. 

\subsection{Optical detection error}
\label{sec:si-optical-detection}
\subsubsection{Choice of the detection parameters}
We use a laser detuning of $\approx -5\MHz$ relative to the measured line center to maximize the ion fluorescence signal while ensuring we do not heat the ion too much during the multiple detection steps, which could lead to a reduction of fluorescence or even complete ion loss \cite{leupold_sustained_2018}. We use a laser intensity of $\sim 100I_\mathrm{sat}$, where the saturation intensity $I_\mathrm{sat}$ is defined as in Ref.~\cite{szwer_high_2009}. This intensity is chosen by observing the signal from the ion at a fixed laser detuning of $-5\MHz$ and finding the point at which the signal saturates, i.e. there is no appreciable increase in the fluorescence rate with the laser power.

We use a camera exposure time of $400\us$ -- as we discuss later, this is the shortest duration at which the optical detection error is much lower than the probability of decay of the metastable level. The camera exposure time is the duration for which the camera collects fluorescence from the ions. However, the total duration of the population detection pulse is slightly longer due to the time it takes for the acquired image to be read out and stored in the buffer before the acquisition of the next image can begin. In our case, this results in a total detection pulse duration of $458.6\us$.

\subsubsection{Detection threshold calibration}
To determine the detection discrimination threshold, we first perform a series of population detection pulses with a single ion in the trap prepared in the \ground level and record the resulting counts. The results from these population detections should all be bright. We then perform the same number of population detection pulses, but this time turning off the $650\nm$ laser. In this case, the population will get shelved in the metastable \dthree level, and the outcomes of the population detection are expected to be dark.

\begin{figure}[!ht]
    \centering
    \includegraphics{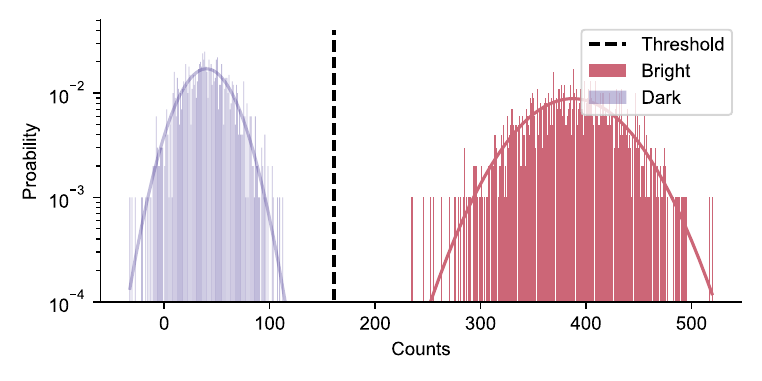}
    \caption{Detection threshold calibration. The detection threshold is obtained by fitting Gaussian curves to the dark and bright histograms and choosing the number of counts that minimises the overlap between the two curves. In the example above, this threshold is $161$ counts. This is the threshold used for the rest of the data in this paper. Each histogram contains $1000$ shots.}
    \label{fig:threshold-calibration}
\end{figure}

The threshold for discriminating between bright and dark is then set by fitting Gaussian functions to the bright and dark distributions and minimizing the overlap between the two fitted curves, as discussed in Ref.~\cite{burrell_high_2010}. An example of such a calibration is shown in Fig.~\ref{fig:threshold-calibration}.

\subsubsection{Optical detection error}
The optical detection error is the error associated with an ion that fluoresces during population detection (i.e. is within the \ground manifold) being measured as dark, an ion that doesn't fluoresce during population detection  (i.e. an ion that starts and remains within \shelf the population detection pulse, or equivalently, the lack of an ion) being recorded as bright. This error can be decreased by increasing the signal generated by a fluorescing ion or, equivalently, by increasing the efficiency of the imaging system used to collect light from the ions and/or the camera exposure time. In this section, we evaluate the quality of this discrimination independently of the decay of the metastable level. We will call the probability to classify an ion in the \ground level as bright $P(b|S_{1/2})$ and the probability to classify an ion in the \shelf level (that remained in the \shelf level during the population detection pulse) to be dark as $P(d|D_{5/2})$.

To evaluate the error of a bright outcome, i.e. $1-P(b|S_{1/2})$, we initialize the ion into $\ket{S_{1/2}, F=2, m_F=0}$ via optical pumping and then perform three consecutive population detection pulses. The error of a bright outcome is given by the probability of a dark result in the second pulse, provided the first and third pulses result in a bright outcome. We repeat this experiment $10^6$ times to accumulate sufficient statistics.

To evaluate the error of a dark outcome, i.e. $1-P(d|D_{5/2})$, we initialize the ion into the $\ket{S_{1/2}, F=2, m_F=0}$ state and then transfer the population to $\ket{D_{5/2}, F=2, m_F=-1}$. We perform three consecutive population detection pulses. The error of a dark outcome is given by the probability of a bright result after the second pulse provided the first and third pulses result in a bright outcome. By demanding that the first and third pulses result in a dark outcome, we are able to decouple the error due to decay of the metastable level (see section \ref{sec:si-metastable-decay}) from the optical detection error.We repeat this experiment $10^6$ times to accumulate sufficient statistics.

\begin{figure}[!ht]
    \centering
    \includegraphics{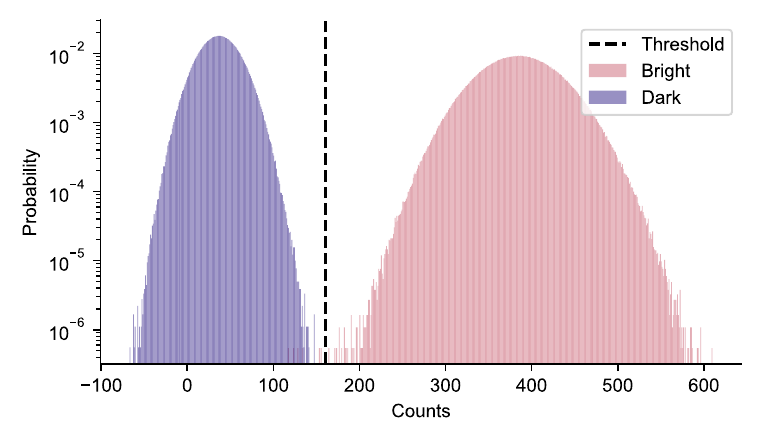}
    \caption{Optical detection error experiment. The histogram shows counts during the second population detection pulse conditioned on correct outcome in the first and third population detection pulses. The optical detection error is estimated by counting the probability for a dark ion to be recorded above the threshold, and for a bright ion to be recorded below the threshold.}
    \label{fig:optical-detection}
\end{figure}

The results from these experiments are shown in Fig.~\ref{fig:optical-detection}. We find a bright state error of $1-P(b|S_{1/2}) = 3.2(28)\times 10^{-6}$ and a dark state error of $1-P(d|D_{5/2}) = 0.0(1.1)\times 10^{-6}$. The slight difference in bright and dark errors is caused by the non-gaussian tail of the bright count distribution (we did not investigate its origin).

\subsection{Metastable level decay probability}
\label{sec:si-metastable-decay}
\begin{figure}[!ht]
    \centering
    \includegraphics{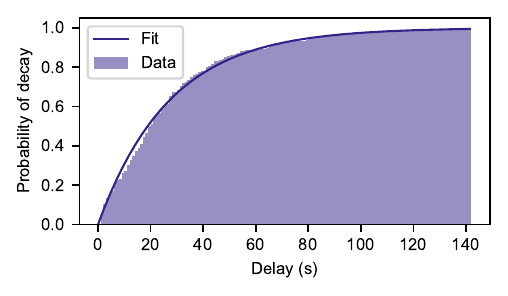}
    \caption{Probability of decay of the metastable level as a function of the delay time between preparation in the metastable level and measurement. The solid line is an exponential fit to the data, $\epsilon_\mathrm{decay}=1-e^{-t_d/\tau}$. The fit corresponds to a decay time of $\tau=27.2(2)\s$.}
    \label{fig:decay-time}
\end{figure}

After the ion is prepared in \shelf, it decays into \ground with probability $\epsilon_\mathrm{decay}=1-e^{-t/\tau}$, where $t$ is the time after preparation and $\tau$ is the lifetime of the \shelf state. To measure $\tau$, we first optically pump the ion into $\ket{S_{1/2}, F=2, m_F=0}$ and then map it to $\ket{D_{5/2}, F=2, m_F=-1}$ followed by a population detection pulse to confirm successful preparation into the metastable level. We then continuously measure the photon count rate from the ion. Once a bright measurement is obtained, the duration for which the ion was dark is recorded, and the procedure is repeated. The results from this experiment are shown in Fig.~\ref{fig:decay-time}. We observe a lifetime of $\tau=27.2(2)\s$. This is slightly lower than the natural lifetime of the \shelf level in \ba of $\tau_n\approx 30.14\s$ \cite{zhang_branching_2020}. We believe this is limited by leakage of the $614\nm$ laser light used to drive the $D_{5/2}\leftrightarrow P_{3/2}$ transition.

We now calculate the probability that an ion originally encoded in the \shelf level is measured as bright in our protocol due to decay, assuming perfect optical detection. This is only relevant for the population detection step $R_3$ used to infer the quantum measurement result. A decay during any of the remaining population detection steps may lead to unnecessarily flagged shots but not an error in the final measurement result.

Let's assume that for a given population detection pulse of length $t_d$, if the ion decays within some time $t_0$ from the start of the pulse, the result would be a bright measurement. However, if the ion decays during the last $(t_d-t_0)$ of the pulse, the result would still be a dark measurement, since it did not spend enough time in the \ground level to be recorded as bright. Now consider the measurement sequence as in Fig.~\ref{fig:spam-protocol}(b), with $R_2$ followed by $R_3$. Assume further that the population transfer step is much shorter than the population detection step, such that $R_2$ and $R_3$ can be considered as occurring in immediate succession. This is a good approximation in our experiment since the average duration of the population transfer pulses is $\approx20\us$, compared to the total detection pulse duration of $t_d=458.6\us$.  In this case, if an ion in \shelf decays during time $t_0$ from the start of $R_2$, the outcome of $R_2$ is bright, so an error flag is raised. Hence, the only way a SPAM error can occur (i.e. the outcome of $R_3$ is bright) without raising an error flag is if the decay occurs during the last $(t_d - t_0)$ of $R_2$, or during the first $t_0$ of $R_3$. Thus, the relevant timescale for decay is $(t_d - t_0) + t_0 = t_d$. Based on the lifetime measurement above, we thus expect a dark ion to be measured as bright during $R_3$ with probability $\epsilon_{\mathrm{decay}} = 1.684(10)\times 10^{-5}$.

\subsection{Total population detection error and SPAM protocol error}
\label{sec:total-error-budget}
The total error for a single population detection can be calculated by combining the optical detection error and the metastable decay probability during the detection pulse. The error for an ion that was prepared into a bright state is then given by the bright state optical detection error i.e. $1-P(b|S_{1/2}) = 3.2(28)\times 10^{-6}$. Conversely, the error in the readout of an ion that was prepared into the dark state is given by $\left(1 - P(dark|D_{5/2}) + \epsilon_\mathrm{d}\times P(b|S_{1/2})\right) = 1.4(6)\times 10^{-5}$. The average population detection error for the two input states is then $8.6(34)\times10^{-6}$.

\section{SPAM success probability}
\label{sec:si-data-rate}
The protocol presented in this paper effectively eliminates errors acquired during standard SPAM procedures at the expense of a reduced success probability. This success probability is determined by the errors in the operations in all steps of the protocol, such as optical pumping and the $S_{1/2}\leftrightarrow D_{5/2}$ population transfer pulses. The measured error probabilities for each of these processes are shown in Table \ref{tab:data-rate}.

In what follows, we calculate the expected fraction of rejected shots at each of the protocol steps. The model includes optical pumping errors and population transfer errors, as well as qubit decay, but ignores the optical readout errors, which are orders of magnitude smaller. The calculated results are shown in able \ref{tab:shot-rejection} and compared with experimental observations. We find that many of the measured values agree well with calculations, and we attribute any differences to system drifts throughout the measurement period.

\begin{table}[h]
    \centering
    \begin{tabular}{|c|c|}
        \hline
        Error type & Error $(\%)$  \\
        \hline
        \hline
         \rule{0pt}{4ex} 
         Optical pumping $(\epsilon_{op})$ & $0.80(6)$ \\[4pt]
         \hline
         \rule{0pt}{4ex}
         $\ket{F=2, m_F=0}\leftrightarrow\ket{F=2, m_F=-1}$ & $1.38(7)$\\[4pt]
        \hline
         \rule{0pt}{4ex}
        $\ket{F=2, m_F=0}\leftrightarrow\ket{F=1, m_F=-1}$ & $4.73(14)$\\[4pt]
        \hline
         \rule{0pt}{4ex}
         $\ket{F=2, m_F=0}\leftrightarrow\ket{F=2, m_F=+1}$ & $3.10(11)$\\[4pt]
        \hline
         \rule{0pt}{4ex}
         $\ket{F=1, m_F=0}\leftrightarrow\ket{F=2, m_F=-1}$ & $1.11(7)$\\[4pt]
        \hline
         \rule{0pt}{4ex}
         $\ket{F=1, m_F=0}\leftrightarrow\ket{F=1, m_F=-1}$ & $0.98(6)$\\[4pt]
         \hline
    \end{tabular}
    \caption{A breakdown of the errors during each of the steps used in the SPAM sequences for all three qubit types. The first row gives the error of preparing into $\ket{S_{1/2}, F=2, m_F=0}$ using optical pumping. This measurement is performed by first performing optical pumping, then doing four consecutive transfer pulses from $\ket{S_{1/2}, F=2, m_F=0}$ to $\ket{D_{1/2}, F=1, m_F=-1}$, $\ket{D_{5/2}, F=1, m_F=+1}$, $\ket{D_{5/2}, F=2, m_F=-1}$, and $\ket{D_{5/2}, F=2, m_F=+1}$, then performing population detection and measuring the probability of a bright outcome. The multiple $1762\nm$ pulses ensure that the error due to population transfer to \ground is negligible, as it is $\sim \epsilon_{1762}^4$ where $\epsilon_{1762}$ is the error of a single pulse. The remaining rows give the errors on the transfer pulses used to map between states in \ground and \shelf. The first state is the state in \ground and the second state is the state in \shelf. The error measurements are performed by first preparing each individual metastable state (using post-selection) and then performing a mapping pulse from that state to \ground and measuring the probability of a dark outcome.}
    \label{tab:data-rate}
\end{table}

\begin{table}[h]
    \centering
    \begin{tabular}{|c|c|c|c|}
        \hline
        Qubit type & State & Shots rejected $(\%)$ & Expected shots rejected $(\%)$  \\
        \hline
        \hline
         \rule{0pt}{4ex} 
         O & $\ket{S_{1/2}, F=2, m_F=0}$ & $10.98$ & $15.0(4)$ \\[4pt]
         \hline
         \rule{0pt}{4ex}
         O & $\ket{D_{5/2}, F=2, m_F=-1}$ & $3.51$ & $3.56(15)$ \\[4pt]
         \hline
         \rule{0pt}{4ex}
         M & $\ket{D_{5/2}, F=2, m_F=-1}$ & $3.41$ & $3.56(15)$ \\[4pt]
         \hline
         \rule{0pt}{4ex}
         M & $\ket{D_{5/2}, F=1, m_F=-1}$ & $8.74$ & $10.26(29)$ \\[4pt]
         \hline
         \rule{0pt}{4ex}
         G & $\ket{S_{1/2}, F=2, m_F=0}$ & $9.21$ & $9.76(27)$ \\[4pt]
         \hline
         \rule{0pt}{4ex}
         G & $\ket{S_{1/2}, F=1, m_F=0}$ & $5.72$ & $5.25(17)$ \\[4pt]
         \hline
    \end{tabular}
    \caption{A comparison of the measured and expected total fraction of rejected shots during the SPAM protocol. We see a good agreement between the two. Any discrepancies are attributed to drifts in system performance between measurements.}
    \label{tab:shot-rejection}
\end{table}

\section{Observable statistics}
\label{sec:superposition-readout}
\newcommand{\prob}[1]{\ensuremath{\text{P}({#1})}}
\newcommand{\condprob}[2]{\ensuremath{\text{P}({#1}\mid{#2})}}
\newcommand{\zm}{\ensuremath{\braket{\mathcal{Z_{\text{meas}}}}}}
\newcommand{\zr}{\ensuremath{\braket{\mathcal{Z}}}}

In our SPAM protocol, the probability of discarding a shot depends in general on the qubit state. As a result, when averaging over multiple experimental shots, incorrect probabilities may be assigned to different measurement outcomes. In this section, we quantify this effect, perform validation experiments, and describe the mitigation strategies.

Consider the problem of measuring the observable $\mathcal{Z} = \ket{0}\bra{0} - \ket{1}\bra{1}$. We can estimate its value using the formula $\zr = \prob{0} - \prob{1}$, where $\prob{0}$ and $\prob{1}$ are the probabilities that the measurements projects the input state into $\ket{0}$ and $\ket{1}$ respectively. In the experiment, we associate those with bright ($b$) and dark ($d$) outcomes of $R_3$ respectively, i.e. ideally $\prob{0} = \prob{b}$ and $\prob{1} = \prob{d}$, such that $\zr = \prob{b} - \prob{d}$.

However, after post-selection, the apparent observable average measured in the experiment is given by $\zm = \condprob{b}{a} - \condprob{d}{a}$, where the probabilities are now conditional on the shot being accepted, which we denote as $a$. Because the probability $\condprob{a}{0}$ for accepting the input state $\ket{0}$ can differ from the probability $\condprob{a}{1}$ for accepting the input state $\ket{1}$, it is possible that $\zm - \zr \neq 0$, leading to a bias in the observable measurement.

We can write the probability to measure a bright outcome as
\begin{equation}
    \prob{b, a} = \sum_{i} \prob{b, a, i} = \sum_{i} \condprob{b, a}{i}\prob{i}.
\end{equation}
where $i \in \{0,1\}$ denotes the qubit state after an ideal projective measurement. Using $\prob{0} + \prob{1} = 1$ we can calculate:
\begin{equation}
    \begin{split}
        \prob{0} &= \frac{\prob{b, a} - \condprob{b, a}{1}}{\condprob{b, a}{0} - \condprob{b, a}{1}},\\
         \prob{1} &= \frac{\prob{b, a} - \condprob{b, a}{0}}{\condprob{b, a}{1} - \condprob{b, a}{0}},\\
        \zr &= \frac{2\prob{b, a} - \condprob{b, a}{1} - \condprob{b, a}{0}}{\condprob{b, a}{0} - \condprob{b, a}{1}}.
    \end{split}
\end{equation}
One method to correct the bias is to measure $\condprob{b, a}{i}$ by preparing each qubit state $\ket{i}$ independently, and then extract the unbiased results from experimental results using the equations above.

Using conditional probability, we can write $\condprob{b, a}{i} = \condprob{b}{a, i}\condprob{a}{i}$. This highlights the two error sources that may bias the observable measured. $\condprob{b}{a, i}$ characterizes the single shot accuracy of the implementation of the protocol. In the ideal case this is $\condprob{b}{a, 0} = 1$ and $\condprob{b}{a, 1} = 0$, but in our system, decay error reduces the latter by  $\epsilon_{\mathrm{decay}} \approx 1.684(10)\times 10^{-5}$, see Sec.~\ref{sec:si-error-budget}. On the other hand, $\condprob{a}{i}$, characterizes the probability of a shot measuring state $\ket{i}$ to be accepted. In our system, imperfect population transfer between $A$ and $B$ creates population transfer errors at the level of $\sim 5\%$, as characterized in Section \ref{sec:si-data-rate}.

\begin{figure}[!ht]
    \centering
    \includegraphics{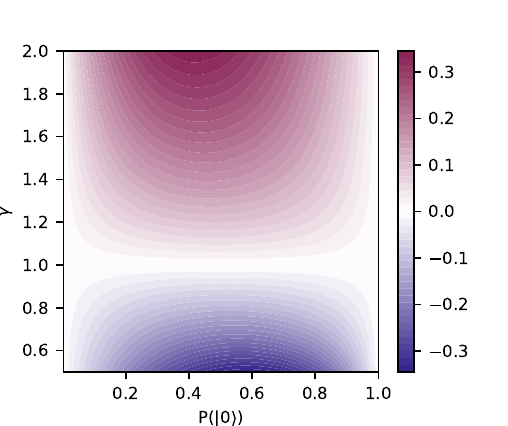}
    \caption{A plot of $\zm - \zr$ with $\gamma \in [\frac{1}{2} , 2]$ and $\prob{0} \in [0, 1]$, with $\zm$ calculated using Eq. \ref{eq:supp-bias}. The bias tends to zero in the cases where $\gamma \to 1$, $\prob{0} \to 0$ or $\prob{0} \to 1$.  }
    \label{fig:supp-asymmetry-sim}
\end{figure}

Considering the decay error to be negligible compared to the error from population transfer pulses, we can simplify the equations above to read
\begin{equation}
    \begin{split}
        \prob{0} &= \frac{\prob{b, a}}{\condprob{a}{0}}\\
        &= \frac{\condprob{b}{a}\prob{a}}{\condprob{a}{0}}, \\
        \prob{1} &= \frac{\condprob{d}{a}\prob{a}}{\condprob{a}{1}}.
    \end{split}
\end{equation}
The apparent observable $\zm$ is then given by
\begin{equation}
    \begin{split}
        \zm &= \condprob{b}{a} - \condprob{d}{a} \\
        &= \frac{\gamma \prob{0} - \prob{1}}{\gamma \prob{0} + \prob{{1}}},
    \end{split}
    \label{eq:supp-bias}
\end{equation}
where $\gamma = \frac{\condprob{a}{0}}{\condprob{a}{1}}$. If $\condprob{a}{0} = \condprob{a}{1}$, we recover $\zr$. In fig. \ref{fig:supp-asymmetry-sim}, we use Eq.~\ref{eq:supp-bias} to plot the bias $\zm - \zr$ for $\gamma \in (\frac{1}{2} , 2)$ and $\prob{0} \in (0, 1)$. 

\begin{figure}[!ht]
    \centering
    \includegraphics[width=0.6\linewidth]{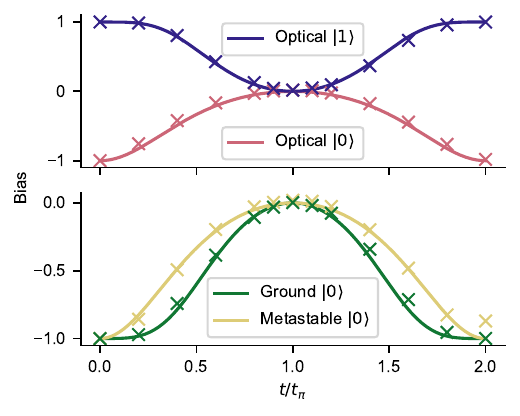}
    \caption{A plot of bias $\zm - \zr $ against the duration duration $t$ of the population transfer pulse. Each curve corresponds to varying the duration of a pulse mapping the corresponding qubit state between the \ground and \shelf manifolds. For the optical qubit, $\ket{0}$ is in the \shelf manifold and $\ket{1}$ is in the \ground manifold. For the ground and metastable qubits, the qubit states are in the same manifold, so the curves for varying the duration of the mapping pulse for $\ket{0}$ and $\ket{1}$ overlap, and we only show the results for $\ket{0}$ for clarity. Measurements are plotted against calculations. To reduce the error in statistics from not faithfully preparing an equal superposition, we alternate the initial qubit state $\ket{0}$ or $\ket{1}$ from which a superposition is created for each shot.}
    \label{fig:asymmetry_infidelity}
\end{figure}

We experimentally measure the observable bias as follows. We start by preparing into $\ket{0}$ or $\ket{1}$ as outlined in Fig.~\ref{fig:spam-protocol} in the main text. Next, we perform a $\pi/2$ rotation to produce an equal superposition state of $\ket{0}$ and $\ket{1}$, where we expect $\zr = 0$. The qubit state is then measured as in Fig.~\ref{fig:spam-protocol} in the main text. We change the duration $t$ of individual population transfer pulses between $A$ and $B$ from its pre-calibrated value of $t_\pi$ to simulate an error in the pulses. The measurement outcome $\zm = \condprob{b}{a} - \condprob{d}{a}$ is then calculated and plotted in Fig. \ref{fig:asymmetry_infidelity}. In the same graph, we also plot the expected functional form of the bias based in Eq.~\ref{eq:supp-bias} and the expected dependence of $\condprob{a}{0}$ and $\condprob{a}{1}$ as tabulated in Tab.~\ref{tab:supp-asymmetry-table}. We find that the measured bias closely matches the model, indicating that data post-processing using Eq.~\ref{eq:supp-bias} can be used to eliminate bias in observable measurements. 

\begin{table}[!ht]
    \centering
    \begin{tabular}{|c|c|c|}
        \hline
        Name & $\condprob{a}{0}$ & $\condprob{a}{1}$ \\
        \hline
        \hline
        \rule{0pt}{4ex} 
        Optical $\ket{0}$ & $\sin^2(t/2t_\pi)$ & 1\\[4pt]
        \hline
        \rule{0pt}{4ex}
        Optical $\ket{1} $ & 1 & $\sin^4(t/2t_\pi)$\\[4pt]
        \hline
        \rule{0pt}{4ex}
        Metastable $\ket{0} $ & $\sin^2(t/2t_\pi)$ & 1\\[4pt]
        \hline
        \rule{0pt}{4ex}
        Ground $\ket{0} $ & $\sin^4(t/2t_\pi)$ & 1\\[4pt]
        \hline
    \end{tabular}
    \caption{The parameters used to simulate the bias from scans in fig. \ref{fig:asymmetry_infidelity}. The entries in Optical $\ket{0}$ and Metastable $\ket{0}$ use $\sin^2(t/2t_\pi)$ since the imperfect mapping will result from just a single pulse that maps \emph{B} (\shelf) $\to$ \emph{A} (\ground). In the case of Optical $\ket{1}$ and Ground $\ket{0}$, there are two pulses, leading to $\sin^4(t/2t_\pi)$. One mapping $A \to B$ before $R2$, and another $B \to A$ before $R3$.}
    \label{tab:supp-asymmetry-table}
\end{table}

\end{document}